\documentclass[iop]{emulateapj}

\usepackage{epsfig}
\usepackage{graphicx}
\usepackage{epstopdf}
\usepackage{amsmath}
\usepackage{natbib}

\begin{document}

\title{Systematic Effects in Interferometric Observations of the CMB Polarization}
\author{Ata Karakci$^{1}$,
 Le Zhang$^{2}$,
 P.~M.~Sutter$^{3,4,5,6}$,
 Emory~F.~Bunn$^{7}$,
 Andrei Korotkov$^{1}$,
 Peter Timbie$^{2}$,
 Gregory~S.~Tucker$^{1}$,
 and Benjamin~D.~Wandelt$^{3,4,5,8}$\\
{~}\\
$^{1}$Department of Physics, Brown University, 182 Hope Street, Providence, RI 02912, USA\\
$^{2}$Department of Physics, University of Wisconsin, Madison, WI 53706, USA\\
$^{3}$ Department of Physics, 1110 W Green Street, University of Illinois at Urbana-Champaign, Urbana, IL 61801, USA\\
$^{4}$ UPMC Univ Paris 06, UMR 7095, Institut d'Astrophysique de Paris, 98 bis, boulevard Arago, 75014 Paris, France\\
$^{5}$ CNRS, UMR 7095, Institut d'Astrophysique de Paris, 98 bis, boulevard Arago, 75014 Paris, France\\
$^{6}$ Center for Cosmology and Astro-Particle Physics, Ohio State University, Columbus, OH 43210, USA\\
$^{7}$ Physics Department, University of Richmond, Richmond, Virginia 23173, USA\\
$^{8}$ Department of Astronomy, University of Illinois at Urbana-Champaign, Urbana, IL 61801, USA
}

\thanks{Email:  ata$\_$karakci@brown.edu}

%\date{\today}
%\maketitle

% Usually omit these for ApJ or MNRAS style files:
%\tableofcontents
%
%\listoffigures
%
%\listoftables

\begin{abstract}

The detection of the primordial $B$-mode spectrum of the polarized cosmic microwave background (CMB) signal may provide a probe of inflation. However, observation of such a faint signal requires excellent control of systematic errors. Interferometry proves to be a promising approach for overcoming such a challenge. In this paper we present a complete simulation pipeline of interferometric observations of CMB polarization, including systematic errors.  We employ two different methods for obtaining the power spectra from mock data produced by simulated observations: the maximum likelihood method and the method of Gibbs sampling. We show that the results from both methods are consistent with each other, as well as, within a factor of 6, with analytical estimates. Several categories of systematic errors are considered: instrumental errors, consisting of antenna gain and antenna coupling errors, and beam errors, consisting of antenna pointing errors, beam cross-polarization and beam shape (and size) errors. In order to recover the tensor-to-scalar ratio, $r$, within a $10\%$ tolerance level, which ensures the experiment is sensitive enough to detect the $B$-signal at $r=0.01$ in the multipole range $28 < \ell < 384$, we find that, for a QUBIC-like experiment, Gaussian-distributed systematic errors must be controlled with precisions of $|g_{rms}| = 0.1$ for antenna gain, $|\epsilon_{rms}| = 5 \times 10^{-4}$ for antenna coupling, $\delta_{rms} \approx 0.7^\circ$ for pointing, $\zeta_{rms} \approx 0.7^\circ$ for beam shape, and $\mu_{rms} = 5 \times 10^{-4}$ for beam cross-polarization.

\begin{center}
\emph{Subject headings}: cosmic background radiation - cosmology:observations - instrumentation:interferometers - methods: data analysis - techniques: polarimetric
\end{center}
\end{abstract}

%Section heading
\section{Introduction}

The Cosmic Microwave Background (CMB) has become one of the most fundamental tools for cosmology. High-precision measurements of the CMB polarization, especially detecting the primordial ``B-mode'' polarization signals~\citep{1997PhRvL..78.2058K}, will represent a major step towards understanding the extremely early universe. These $B$-modes are generated by primordial gravitational waves. A detection of these signals would probe the epoch of inflation and place an important constraint on the inflationary energy scale~\citep{2002ARA&A..40..171H}.   In addition, the secondary $B$-modes induced by gravitational lensing encode information about the distribution of dark matter. However, the $B$-mode signals are expected to be extremely small and current experiments can only place upper limits~\citep{Hinshaw:2012fq} on the tensor-to-scalar ratio; the quest for the $B$-modes is a tremendous experimental challenge.
  
Due to the weakness of the $B$-mode signals -- the largest signal of the primordial $B$-modes is predicted to be less than $0.1\mu$K -- exquisite systematic error control is crucial for detecting and characterizing them. Compared to imaging systems, interferometers offer certain advantages for controlling systematic effects because: (1) an interferometer does not require rapid chopping and scanning~\citep{2006NewAR..50..999T} and, with simple optics, interferometric beam patterns have extremely low sidelobes and can be well understood; (2) interferometers are insensitive to any uniform sky brightness or fluctuations in atmospheric emissions on scales larger than the beam width; (3) without differencing the signal from separate detectors, interferometers measure the Stokes parameters directly and inherently avoid the leakage from temperature into polarization~\citep{Bunn:2006nh} caused by mismatched beams and pointing errors, which are serious problems for $B$-mode detection with imaging experiments~\citep{Hu:2002vu,Su:2010wa,Miller:2008zi,O'Dea:2006di,Shimon:2007au,Yadav:2009za,2010ApJ...711.1141T}; (4) for observations of small patches of sky, $E$-$B$ mode separation would be cleaner in the Fourier domain for interferometric data than in real-space; and (5) with the use of redundant baselines, systematic errors can be averaged out. In addition, they offer a straightforward way to determine the angular power spectrum since the output of an interferometer is the {\it visibility}, that is, the Fourier transform of the sky intensity weighting by the response of the antennas.

Interferometers have proved to be powerful tools for studying the CMB temperature and polarization power spectra. In fact, DASI~\citep{2002Natur.420..772K} was the first instrument to detect the faint CMB polarization anisotropies.  Pioneering attempts to measure the CMB temperature anisotropy with interferometers were made in the 1980s~\citep{1980ApJ...240L..79M,1984Sci...225...23F,1984ApJ...284..479K,1988Natur.331..146P,1988RScI...59..914T}. Several groups have successfully detected the  CMB anisotropies. The CAT telescope was the first interferometer to actually detect structures in the CMB~\citep{1995MNRAS.274..861O,1996ApJ...461L...1S,1999MNRAS.308.1173B}. CBI~\citep{2003ApJ...591..556P} and VSA~\citep{2004MNRAS.353..732D,2003MNRAS.341L..23G} have detected the CMB temperature and polarization angular power spectra down to sub-degree scales. In the next few years, the QUBIC instrument~\citep{2011APh....34..705Q} based on the novel concept of bolometric interferometry is expected to constrain the tensor-to-scalar ratio to $0.01$ at the $90\%$ confidence level, with 1-year of observing.

On the theory side, the formalism for analyzing  interferometric CMB data has been well-developed~\citep{1999ApJ...514...12W,2002MNRAS.334..569H,2003ApJ...589...67P,2003ApJ...591..575M,2006NewAR..50..951M,1996MNRAS.283.1133H}. A pioneering study of systematic effects for interferometers based on an analytic approach has been performed by~\citet{Bunn:2006nh}. However, this approach is of course only a first-order approximation for assessing systematics, since many important effects, such as the configuration of the array, instrumental noise, and the sampling variance due to finite sky coverage and incomplete $uv$-coverage, are not taken into account. Any actual experiment therefore will naturally require a complete simulation to assess exactly how systematic effects bias the power spectrum recovery.  In this concern, ~\cite{Zhang2012} have presented a simulation pipeline to assess the systematic errors, mainly focusing on pointing errors. With a full maximum likelihood (ML) analysis of mock data, the simulation agrees with the analytical estimates and finds that, for QUBIC-like experiments, the Gaussian-distributed pointing errors have to be controlled to the sub-degree level to avoid contaminating the primordial $B$-modes with $r\leq0.01$.  

Nevertheless, a comprehensive and complete analysis of various systematic errors on CMB power spectrum measurements has not been undertaken so far. In this paper, therefore, we perform a detailed study to completely diagnose the most serious systematic effects including gain errors, cross-talk, cross-polarization, beam shape errors and pointing errors, on the entire set of CMB temperature and polarization power spectra. In order to assess the effect of the systematic errors on $B$-mode detection and set allowable tolerance levels for those errors, we perform simulations for a specific interferometric observation with an antenna configuration similar to the QUBIC instrument.  We also extend the analytical expressions~\citep{Bunn:2006nh} for characterizing systematic effects on the full CMB power spectra.

 For verifying the power spectrum analysis, we employ two completely independent codes based on the Gibbs sampling algorithm and the maximum likelihood technique. The use of Gibbs-sampling based Bayesian inference to interferometric CMB observations has been successfully demonstrated by~\cite{Sutter:2011uv} and \cite{Karakci2012}.  It allows extraction of the underlying CMB power spectra and reconstruction of the pure CMB signals simultaneously, with a much lower computational complexity in contrast to the traditional maximum likelihood technique~\citep{2002MNRAS.334..569H}. 

In this paper, for given input CMB angular power spectra, we simulate the observed Stokes visibilities in the flat-sky approximation. We believe that the flat-sky simulations are sufficiently accurate for the study of systematic errors. First, in our simulation, we assume single pointing observations with $5^\circ$ beam width, corresponding to a sky coverage fraction of $f_{sky} =0.37\%$. This sky patch is small enough to permit the use of the flat-sky approximation. Second, all the data analysis processes are established using the flat-sky approximation while the mock visibility data are also simulated using this approximation.  Therefore, a self-consistent analysis is performed. However, when using a patch cut from the projection of spherical sky onto a flat image as an ``input'' map, one should take into account the contamination~\citep{Bunn:2010kf} of ``ambiguous'' modes arising from incomplete sky coverage and thus requires an appropriate data analysis method to apply for this situation.  

This paper is organized as follows. In Sec.~2, we briefly summarize the effects of a variety of systematic errors on interferometric CMB observations and describe the analytical method for estimating those errors.  In Sec.~3, we describe the simulations interferometric visibilities that include systematic errors. In Sec.~4, we review the data analysis methods used in this paper, including the Gibbs sampling technique and the maximum likelihood approach. In Sec.~5, we assess the systematic effects on the CMB power spectra. Finally, a discussion and summary are given in Sec.~6. The appendix contains the complete analytical expressions for the systematic effects on the full CMB power spectra.

%Due to the fact that typically a small patch of the sky is viewed by an interferometer, one can use the ``flat-sky'' approximation to estimate the CMB power spectra from observed visibilities by two-dimensional Fourier transform analysis instead of spherical harmonic expansions. The formalism of CMB visibilities related to power spectra in a curved-sky has been studied by~\cite{2007ApJ...655...21B} and~\cite{2001PhRvD..63l3001N}.The accuracy on recovered CMB power spectra in the flat-sky approximation against with the curved-sky treatment are well determined. If one want to obtain a finner resolution in $\ell$-space and reduce the sampling variance, one has to survey larger areas of the sky~\citep{1999ApJ...514...12W,1996MNRAS.283.1133H} by using antennas with wide field-of-view such as a dipole or cylinder telescope~\citep{2006astro.ph..6104P}, or by ``mosaicing'' several different small patches of the sky together instead of a single sky pointing scanning. 

\section{Systematics}

\subsection{Instrument Errors and Beam Errors}

In a polarimetric experiment, the Stokes parameters $I, Q, U$ and $V$ can be obtained by using either linear or circular polarizers. For a given baseline $ {\mathbf u}_{jk} =  {\mathbf x}_{k} -  {\mathbf x}_{j}$, $ {\mathbf x}_{k}$ being the position vector of the $k^{th}$ antenna, the visibilities can be written as a $2 \times 2$ matrix $\mathbf{V}_{jk}$ ~\citep{Bunn:2006nh};

\begin{equation}\label{eq:vz}
\mathbf{V}_{jk} = \int {d^2 \hat{\mathbf r} ~ \mathbf{A}_{k}(\hat{\mathbf r}) ~ \mathbf{R} \cdot \mathbf{S} \cdot \mathbf{R}^{-1} \mathbf{A}_{j}^{\dagger}(\hat{\mathbf r}) ~ e^{-2\pi i {\mathbf u}_{jk}\cdot \hat{\mathbf r} } },
\end{equation}
where the $2 \times 2$ matrix $\mathbf{A}_{k}(\hat{\mathbf r})$ is the antenna pattern and

\begin{equation} \label{eq:stok} 
\mathbf S = \left( \begin{array}{cc}
I + Q &  U + iV \\
U - iV & I - Q  \end{array} \right).
\end{equation}
For a \emph{linear experiment}, $\mathbf{R}$ is the identity matrix and for a \emph{circular experiment},

\begin{equation} \label{eq:rot} 
\mathbf{R}_{(circ)} = {1 \over \sqrt{2}} \left( \begin{array}{cc}
1 &  i \\
1 & -i  \end{array} \right).
\end{equation}

Various systematic errors can be modeled in the definition of the antenna pattern as follows ~\citep{Bunn:2006nh}

\begin{equation} \label{eq:errs} 
\mathbf{A}_{k}(\hat{\mathbf r}) = \mathbf{J}_{k} \cdot \mathbf{R} \cdot \mathbf{A}^{k}_{s}(\hat{\mathbf r}) \cdot \mathbf{R}^{-1}
\end{equation}
where the \emph{Jones matrix} $\mathbf{J}_{k}$ represents the instrumental errors, such as gain errors and antenna couplings. The matrix $\mathbf{A}^{k}_{s}$ is the antenna pattern that models the beam errors, such as pointing errors, beam shape errors and cross-polarization. In an ideal experiment $\mathbf{J}_{k}  = \mathbb{I}$, where $\mathbb{I}$ is the identity matrix, and the antenna pattern is given as $\mathbf{A}^{k}_{s}(\hat{\mathbf r}) = A(\hat{\mathbf r}) \mathbb{I}$, where $A(\hat{\mathbf r})$ a circular Gaussian function.

In this paper we will consider only two types of instrumental errors; antenna gain, parametrized by $g_1^k$ and $g_2^k$, and couplings, parametrized by $\epsilon_1^k$ and $\epsilon_2^k$.  The coupling errors are caused by mixing of the two orthogonally polarized signals in the system. To account for the phase delays, the parameters $g$ and $\epsilon$ are given as complex numbers. The Jones matrix for the $k^{th}$ antenna can be written as ~\citep{Bunn:2006nh}

\begin{equation} \label{eq:jon} 
\mathbf{J}_k = \left( \begin{array}{cc}
1 + g_1^k &  \epsilon_1^k \\
\epsilon_2^k & 1 + g_2^k  \end{array} \right).
\end{equation}

For the beam errors, we will consider that each antenna has a slightly different beam width, ellipticity (beam shape errors) and beam center (pointing errors), as well as a cross-polar antenna response described by off-diagonal entries in the antenna pattern matrix ~\citep{Bunn:2006nh};

\begin{equation} \label{eq:ant} 
\mathbf{A}_s^k = A_0^k(\rho, \phi) \left( \begin{array}{cc}
 1 +  {1 \over 2} \mu_k {\rho^2 \over \sigma^2} \cos 2\phi &  {1 \over 2} \mu_k {\rho^2 \over \sigma^2} \sin 2\phi \\
  {1 \over 2} \mu_k {\rho^2 \over \sigma^2} \sin 2\phi & 1 -  {1 \over 2} \mu_k {\rho^2 \over \sigma^2} \cos 2\phi\end{array} \right).
\end{equation}
where $A_0^k(\rho, \phi)$ is an elliptical Gaussian function written in polar coordinates $(\rho, \phi)$, $\sigma$ is the width of the ideal beam and $\mu_k$ is the cross-polarization parameter of the $k^{th}$ antenna. This particular form of the cross-polarization occurs, with $\mu_k = \sigma^2 / 2$, when the curved sky patch is projected onto a plane.

\subsection{Control Levels}

The effect of errors on the power spectra can be described by the root-mean-square difference between the actual spectrum, $C^{XY}_{actual} $, which is recovered from the data of an experiment with systematic errors, and the ideal spectrum, $C^{XY}_{ideal} $, which would have been recovered from the data of an experiment with no systematic errors;

\begin{equation} \label{eq:diff} 
\Delta C^{XY} = \left < \left( C^{XY}_{actual} - C^{XY}_{ideal} \right)^2 \right >^{1/2}
\end{equation}
where $X, Y = \{ T, E, B \}$.

The strength of the effect of systematics can be quantified by a tolerance parameter $\alpha^{XY}$ defined by~\citep{O'Dea:2006di, Miller:2008zi, Zhang2012}

\begin{equation} \label{eq:alph} 
\alpha^{XY} = {\Delta C^{XY} \over \sigma_{stat}^{XY}}
\end{equation}
where $\sigma_{stat}^{XY}$ is the statistical 1-$\sigma$ error in $XY$-spectrum of the ideal experiment with no systematic errors.

The main interest in a $B$-mode experiment is the tensor-to-scalar ratio $r$ which can be estimated as~\citep{O'Dea:2006di}

\begin{equation} \label{eq:rval} 
r  = { \sum_b \partial_r C_b^{BB} (C_b^{BB} - C_{b,lens}^{BB})/ (\sigma_{b, stat}^{BB})^2 \over \sum_b ( {\partial_r C_b^{BB} / \sigma_{b, stat}^{BB} } )^2 }
\end{equation}
where $b$ denotes the power band,  $ C_{b,lens}^{BB}$ is the $B$-mode spectrum due to weak gravitational lensing and $C_b^{BB}$ depends linearly on $r$ through the amplitude of the primordial $B$-modes. The tolerance parameter of $r$ is given by $\alpha_r = \Delta r / \sigma_r$~\citep{O'Dea:2006di};

\begin{subequations} \label{eq:alfr} 
\begin{align}
\Delta r  &= { \sum_b  \alpha_b^{BB} (\partial_r C_b^{BB} / \sigma_{b, stat}^{BB})  \over \sum_b ( {\partial_r C_b^{BB} / \sigma_{b, stat}^{BB} } )^2 }, \label{deltr} \\
\sigma_r  &= \left ( \sum_b ( {\partial_r C_b^{BB} / \sigma_{b, stat}^{BB} } )^2 \right )^{-1/2}. \label{sigmr}
\end{align}
\end{subequations}
 For good control of systematics, the value of $\alpha_r$ is required to stay below a determined tolerance limit. 

\subsection{Analytical Estimations}

Analytical estimations of the effect of systematic errors on the polarization power spectra are extensively examined in ~\citet{Bunn:2006nh}. Defining a vector of visibilities $ {\mathbf v} = (V_I, V_Q, V_U)$ corresponding to a single baseline ${\mathbf u}$ pointing in the $x$ direction, for an ideal experiment, we can write

\begin{subequations} \label{eq:vizer} 
\begin{align}
\left< |V_I|^2 \right> & =  C^{TT}_{\ell = 2\pi u}, \label{vizer3} \\
\left< |V_Q|^2 \right> & = C^{EE}_{\ell = 2\pi u} \overline{c^2} + C^{BB}_{\ell = 2\pi u} \overline{s^2}, \label{vizer1} \\
\left< |V_U|^2 \right> & = C^{EE}_{\ell = 2\pi u} \overline{s^2} + C^{BB}_{\ell = 2\pi u} \overline{c^2}, \label{vizer2} \\
\left< V_Q V^{*}_U \right>  & =  C^{EB}_{\ell = 2\pi u} (\overline{c^2} -  \overline{s^2}), \label{vizer6} \\
\left< V_I V^{*}_Q \right>  & =  C^{TE}_{\ell = 2\pi u} \overline{c}, \label{vizer4} \\
\left< V_I V^{*}_U \right>  & =  C^{TB}_{\ell = 2\pi u} \overline{c}. \label{vizer5}
\end{align}
\end{subequations}
where $\overline{c^2}$, $\overline{s^2}$ and $\overline{c}$ are averages of $\cos^2(2\phi)$, $\sin^2(2\phi)$ and $\cos(2\phi)$ over the beam patterns:

\begin{equation} \label{eq:sbar}
\overline{s^2} = {\int |\widetilde{A^2}( {\mathbf k} - 2\pi { {\mathbf u}})|^2 \sin^2(2\phi) d^2 {\mathbf k} \over \int  |\widetilde{A^2}( {\mathbf k})|^2 d^2 {\mathbf k}} = 1 - \overline{c^2},
\end{equation}
where $\widetilde{A^2}$ is the Fourier transform of the ideal beam pattern squared. The unbiased estimator for $C^{XY} = \left < \hat C^{XY} \right > $ is obtained as

\begin{equation} \label{eq:estm} 
\hat C^{XY} = {\mathbf v}^{\dagger} \cdot \mathbf{N}_{XY} \cdot {\mathbf v}
\end{equation}
where $\mathbf{N}_{XY}$ is a $3 \times 3$ matrix involving $\overline{s^2}$ and $\overline{c}$ (see Appendix). For a baseline pointing in an arbitrary direction the analysis is done in a rotated coordinate system:

\begin{equation} \label{eq:rotos} 
 {\mathbf v}_{rot} = \left( \begin{array}{ccc}
1 &  0 & 0 \\
0 & \cos 2\theta & \sin 2\theta \\
0 & -\sin 2\theta & \cos 2\theta \end{array} \right)  {\mathbf v},
\end{equation}
$\theta$ being the angle between ${\mathbf u}$ and the $x$-axis.

The effect of errors on visibilities can be described, to first order, by $ {\mathbf v}_{actual} =  {\mathbf v}_{ideal} + {\delta {\mathbf v}}$. Combining $ {\mathbf v}_{ideal}$ and ${\delta {\mathbf v}}$ into a 6-dimensional vector $ {\mathbf w} = ( {\mathbf v}, \delta {\mathbf  v})$, we can write the first order approximation as ~\citep{Bunn:2006nh}

\begin{equation} \label{eq:delce}
(\Delta \hat C^{XY}_{rms})^2 = Tr [ ({\mathcal{N}}_{XY} \cdot {\mathcal{M}}_w)^2 ] +  (Tr [{\mathcal{N}}_{XY} \cdot {\mathcal{M}}_w])^2,
\end{equation}
where ${\mathcal{M}}_w = \left<  {\mathbf w} \cdot  {\mathbf w} ^{\dagger} \right >$ is the covariance matrix of $ {\mathbf w}$ and

\begin{equation} \label{eq:curvyN} 
 {\mathcal{N}}_{XY} = \left( \begin{array}{cc}
0 & \mathbf{N}_{XY} \\
\mathbf{N}_{XY} & 0 \end{array} \right).
\end{equation}
The error on a particular band power is, then, given as an expansion in terms of ideal power spectra:

\begin{equation} \label{eq:expan}
(\Delta \hat C^{XY}_{rms, b})^2 = p^2_{rms} \sum_{I,J} \kappa^2_{XY, I, J}C_b^I C_b^J
\end{equation}
where $p$ is the parameter that characterizes the error, such as gain, $g$, coupling, $\epsilon$, or cross-polarization, $\mu$, and $I, J = \{TT, TE, EE, BB \}.$ This expression is valid for a single baseline. For a system with $n_b$ baselines in band $b$, $\Delta \hat C^{XY}_{rms, b}$ must be normalized by $1 / \sqrt{n_b}$, assuming there is no correlation between error parameters of different baselines. Analytical estimations of the coefficients $ \kappa^2_{XY, I, J}$ for various systematic errors are presented in the Appendix.

\section{Simulations}

The input $I, Q$ and $U$ maps are constructed over 30-degree square patches with $64$ pixels per side as described in \citet{Karakci2012} with the cosmological parameters consistent with the 7-year results of WMAP~\citep{Larsonal2011, Komatsal2011}. The tensor-to-scalar ratio is taken to be $r = 0.01$.  The angular resolution of the signal maps is 28 arcminutes, corresponding to a maximum available multipole of $\ell_{max} = 384$. The ideal primary beam pattern, $A(\hat{\mathbf r})$, is modeled as a Gaussian with peak value of unity and standard deviation of $\sigma = 5^\circ$, which drops to the value of $10^{-2}$ at the edges of the patch, reducing the edge-effects caused by the periodic boundary conditions of the fast Fourier transformations. Although the patch size is too large for the flat-sky approximation, the width of the primary beam is small enough to employ the approximation. 

The interferometer configuration is  a close-packed square array of 400 antennas with diameters of 7.89$\lambda$. The observation frequency is 150 GHz with a 10-GHz bandwidth. This configuration is similar to the QUBIC design~\citep{2011APh....34..705Q}. With this frequency and antenna radius, the minimum available multipole is $\ell_{min} = 28$.  The baselines are uniformly rotated in the $uv$-plane over a period of 12 hours while observing the same sky patch. The resulting interferometer pattern is shown in Figure 1.

\begin{figure} \label{fig:inter}
\begin{center}
    \leavevmode
    \includegraphics[trim = 2.2cm 0.5cm 2.5cm 1.4cm, clip=true, width=8cm]{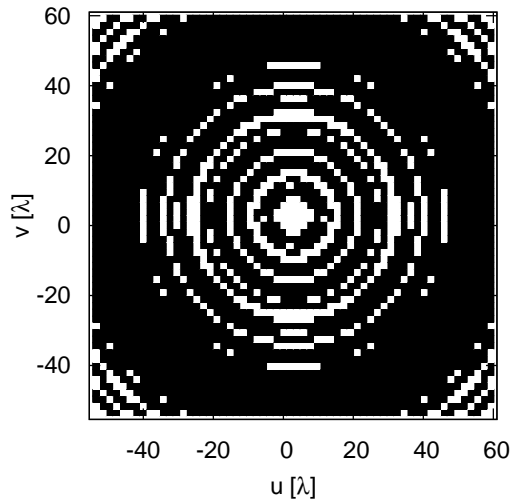}
    \caption{Interferometer pattern created over an observation period of 12 hours by $20 \times 20$ close-packed array of antennas of radius $7.89 \lambda$.}
\end{center}
\end{figure}

The noise at each pixel for the temperature data is obtained from the total observation time that all baselines spend in the pixel. The noise covariance for the baseline ${\mathbf u}_{kj}$ is given as \citep{1999ApJ...514...12W}

\begin{equation} \label{eq:nocvar}
C^{kj}_N = \left({\lambda^2 T_{sys} \over \eta^A A_D} \right)^2 \left({1 \over \Delta_\nu t_a \bar n} \right) \delta_{kj}
\end{equation}
where $T_{sys}$ is the system temperature, $\lambda$ is the observation wavelength, $\eta^A$ is the aperture efficiency, $\Delta_\nu$ is the bandwidth, $\bar n$ is the number of baselines with the same baseline vector, and $t_a$ is the integration time. The noise value is normalized by a constant to have an rms noise level of $0.015 \mu K$ per visibility, yielding an average overall signal-to-noise ratio of about 5 for the $Q$ and $U$ maps.

The systematic errors are introduced by calculating the visibilities in each pixel according to Eq.~\ref{eq:vz}. Each antenna has random error parameters for gain, coupling, pointing, beam shape, and cross-polarization errors drawn from Gaussian distributions with rms values of $|g_{rms}| = 0.1$, $|\epsilon_{rms}| = 5 \times 10^{-4}$, $\delta_{rms} = 0.1 \sigma \approx 0.7^\circ$, $\zeta_{rms} = 0.1 \sigma \approx 0.7^\circ$, and $\mu_{rms} = 5 \times 10^{-4}$, respectively. Here $\delta$ is the offset of the beam centers of the antennas and $\zeta$ is the deviation in the beam width along the principal axes of the elliptical beams. As the baseline rotates, the beam patterns of the corresponding antennas get rotated as well. Whenever a baseline crosses a new pixel, the visibility within the pixel, given by Eq.~\ref{eq:vz}, is calculated again with the rotated beam patterns. The data in a given pixel is taken as the average of all the visibilities calculated in that pixel.

In a circular experiment, the Stokes variables $Q$ and $U$ can be simultaneously obtained for the same baseline. Thus, for a circular experiment, $p_{circ}^{Q} = p_{circ}^{U}$. However, for a linear experiment, direct measurement of $Q$ requires perfect cancellation of the much larger $I$ contribution in Eq.~\ref{eq:stok}. Practically, a linear experiment only measures $U$. Since $U \to Q$ under a $45^\circ$ clockwise rotation, $Q$ can be measured by measuring $U$ with $45^\circ$-rotated linear polarizers. Since $Q$ and $U$ are not measured simultaneously by the same baseline, in general, the error parameters $p_{lin}^Q$ and $p_{lin}^U$ are treated as the distinct parameters in a linear experiment, i.e., $p_{lin}^Q \neq p_{lin}^U$. To simulate this, we calculate $V_U$ with a set of error parameters, $p_{lin}^U$. Then $Q$ and $U$ in Eq.~\ref{eq:stok} are replaced by $-U$ and $Q$, respectively, and $V_U$ is calculated again with a different set of parameters, $p_{lin}^Q$, to obtain $V_Q$. The simulation requires 4.5 CPU-hours for the circular experiment and 13.5 CPU-hours for the linear experiment.

\section{Analysis Methods}

\subsection{Maximum Likelihood Analysis}

The scheme for the maximum likelihood (ML) analysis of CMB power spectra from interferometric visibility measurements is presented in~\cite{Hobson:2002zd,Park:2002ka,Zhang2012}, which we briefly summarize here.  The ML estimator of the power spectrum has many desirable features~\citep{Bond:1998zw,GVK514310375} and has been widely applied in CMB cosmology~\citep{Bond:1998zw,1997ApJ...480....6B,Hobson:2002zd}.

In practice, we divide the total $\ell$-range into $N_b$ spectral bands, each of bin-width $\Delta \ell$. The power spectrum $C_\ell$ thus can be parametrized as flat band-powers $\overline{\mathcal{C}_b} (b=1,\dotsc,N_b)$ over $\Delta \ell$ to evaluate the likelihood function~\citep{1997ApJ...480....6B,Bond:1998zw,1996ApJ...464L..11G,1999ApJ...514...12W}. In each of the band-powers, we assume $\ell(\ell+1)C_\ell$ to be a constant value to characterize the averaged $C_\ell$ over $\Delta \ell$ and has $\overline{\mathcal{C}_b}\equiv 2\pi|{\bf u}_b|^2S(|{\bf u}_b|)$ as the flat-sky approximation~\citep{1999ApJ...514...12W}. 

In our case, the CMB signals and the instrumental noise are assumed to be Gaussian random fields. Therefore, for a given set of CMB band-power parameters $\{\overline{\mathcal{C}_b^{TT}},\overline{\mathcal{C}_b^{EE}},\overline{\mathcal{C}_b^{BB}},\overline{\mathcal{C}_b^{TE}},\overline{\mathcal{C}_b^{TB}},\overline{\mathcal{C}_b^{EB}}\}$, the signal covariance matrices can be written as  
\begin{equation}
C^{ij}_{ZZ'} = \sum_{b=1}^{N_b} \sum_{X, Y} \overline{\mathcal{C}_b^{XY}}  \int _{|{\bf u}_{b1}|}^{|{\bf u}_{b2}|}\, \frac{1}{2\pi}\frac{dw}{w} \times  W^{i,j}_{ZZ'XY}(w) \, ,
\label{eq:eqczb}
\end{equation}
where we introduced the so-called {\em window functions} $W^{ij}_{ZZ'XY}$ given by 
 
\begin{equation} 
W_{ZZ'XY}^{ij}(|{\bf w}|) = \int_0^{2\pi}d\phi_{\bf w}\, \omega_{ZX} \omega_{Z'Y}   \tilde{A}({\bf u}_i-{\bf w})\tilde{A}^*({\bf u}_j-{\bf w}) \, ,
\label{eq:wf} 
\end{equation}
where $Z,Z' = \{I,Q,U\}$ and $X,Y =\{T,E,B\}$ with $\omega_{IT}=1$, $\omega_{UE} = \sin 2\phi_{\bf w}$, $\omega_{UB} = \cos 2\phi_{\bf w}$, $\omega_{QE} = \cos 2\phi_{\bf w}$, $\omega_{QB} = -\sin 2\phi_{\bf w}$ and otherwise zero.
 
Due to the fact that the window functions $W_{ZZ'XY}^{ij}(|{\bf w}|)$ are independent of $\overline{\mathcal{C}_b}$, the integrals of the window functions over $w$ in Eq.~\ref{eq:eqczb} only have to be calculated once before evaluating the covariance matrices. Additionally, if the primary beam pattern $A({\bf x})$ is Gaussian, the window functions can be expressed analytically (see details in ~\cite{2002MNRAS.334..569H,2003ApJ...589...67P,Zhang2012}).

We evaluate the likelihood function by varying the CMB band-powers using the above parametrization. Following~\cite{Hobson:2002zd,2003ApJ...589...67P,2003ApJ...591..575M,Zhang2012}, the logarithm of the likelihood function for interferometric observations is given by
\begin{equation} 
\ln\mathcal{L}(\{\overline{\mathcal{C}_b}\}) = n\log\pi-\log|C_V+C_N| - {\bf d}^\dagger_V(C_V+C_N)^{-1}{\bf d}_V  \, , 
\label{eq:loglike} 
\end{equation}
where $C_V$ is the predicted signal covariance matrix and $C_N$ is the instrumental noise covariance matrix, ${\bf d}_V$ is the observed visibility data vector constructed by ${\bf d}_V \equiv (\dotsb;V_I({\bf u}_i),V_Q({\bf u}_i),V_U({\bf u}_i);\dotsb) (i=1,\dotsc,n)$ where $i$ denotes the visibility data contributed from the pure CMB signals and the instrument noise at the $i$-th pixel in the $uv$-plane and we have a total of $n$ data points.

As mentioned by~\cite{2002MNRAS.334..569H} and references therein, the combination of the {\em sparse matrix conjugate-gradient} technique and {\em Powell's directional-set} method give a sophisticated and optimized numerical algorithm for maximizing the likelihood function to find the ``best-fitted'' CMB power spectrum.  With an appropriate initial guess to start iteration, independent line-maximization is performed for each band-power parameter in turn, while fixing the others. Typically, this process requires a few iterations, of order $N_b^2$, to achieve the maximum-likelihood solution. For about 4000 visibilities in a QUBIC-like observation, the maximum-likelihood solution of $6\times6$ CMB band-powers can be obtained in around 20 CPU-hours. 

Assuming the likelihood function near its peak can be well-approximated by a  Gaussian, the confidence level of the derived maximum-likelihood CMB power spectrum is given by the inverse of the curvature (or Hessian) matrix at the peak. The Hessian matrix  is the matrix of second derivatives of the log-likelihood function with respect to the parameters. This matrix is easily evaluated numerically by performing second differences along each parameter direction. The square roots of the diagonal elements of the inverse of the Hessian matrix give the standard error on each band-power. This procedure requires only about 30 CPU-mins for $\sim4000$ visibilities.

\subsection{Gibbs Sampling Method}

As discussed in \citet{Karakci2012},  the method of Gibbs sampling has been applied to interferometric observations of the polarized CMB signal in order to recover both the input signal and the power spectra.

The CMB signal is described as a $3n_p$ dimensional vector, $ {\mathbf s}$, of the Fourier transform of the pixelated signal maps of $n_p$ pixels; $ {\mathbf s} = (..., \tilde{T}_i, \tilde{E}_i, \tilde{B}_i, ...); ~ i = 0, ..., n_p - 1$. %Similarly, the visibility data are described as a $3n_p$ dimensional vector, $ {\mathbf d}_V$.

The Gibbs sampling method is employed to sample the signal, $ {\mathbf s}$, and the signal covariance, $\mathbf S = \left <  {\mathbf s} ~  {\mathbf s} ^{\dagger} \right >$, from the joint distribution $P(\mathbf{S}, ~{\mathbf s}, ~{\mathbf d}_V)$ by successively sampling from the conditional distributions in an iterative fashion \citep{Larson2007, Karakci2012}:

\begin{subequations} \label{eq:sampl} 
\begin{align}
 {\mathbf s}^{a+1} & \leftarrow  P( {\mathbf s} ~|~ \mathbf{S}^a, ~ {\mathbf d}_V) \label{sampl3} \\
\mathbf S^{a+1} & \leftarrow  P( \mathbf{S} ~|~  {\mathbf s} ^{a+1}). \label{sampl4}
\end{align}
\end{subequations}
After a ``burn-in" phase, the stationary distribution of the Markov chain is reached and the samples approximate to being samples from the joint distribution.

To determine that the stationary distribution of the Markov chain has been reached, the Gelman-Rubin (GR) statistic is employed \citep{Gelman92, Sutter:2011uv, Karakci2012}. For multiple instances of chains, when the ratio of the variance within each chain to the variance among chains drops to a value below a given tolerance, the convergence is said to be attained. The convergence of the Gibbs sampling is reached roughly in 30 CPU-hours.

\section{Results}

\subsection{Power Spectra}

\begin{figure*} \label{fig:spect}
  \begin{center}$
    \leavevmode
    \begin{array}{c@{\hspace{1.5cm}}c}
     \includegraphics[trim = 1mm 1mm 1mm 1mm, width=7.5cm]{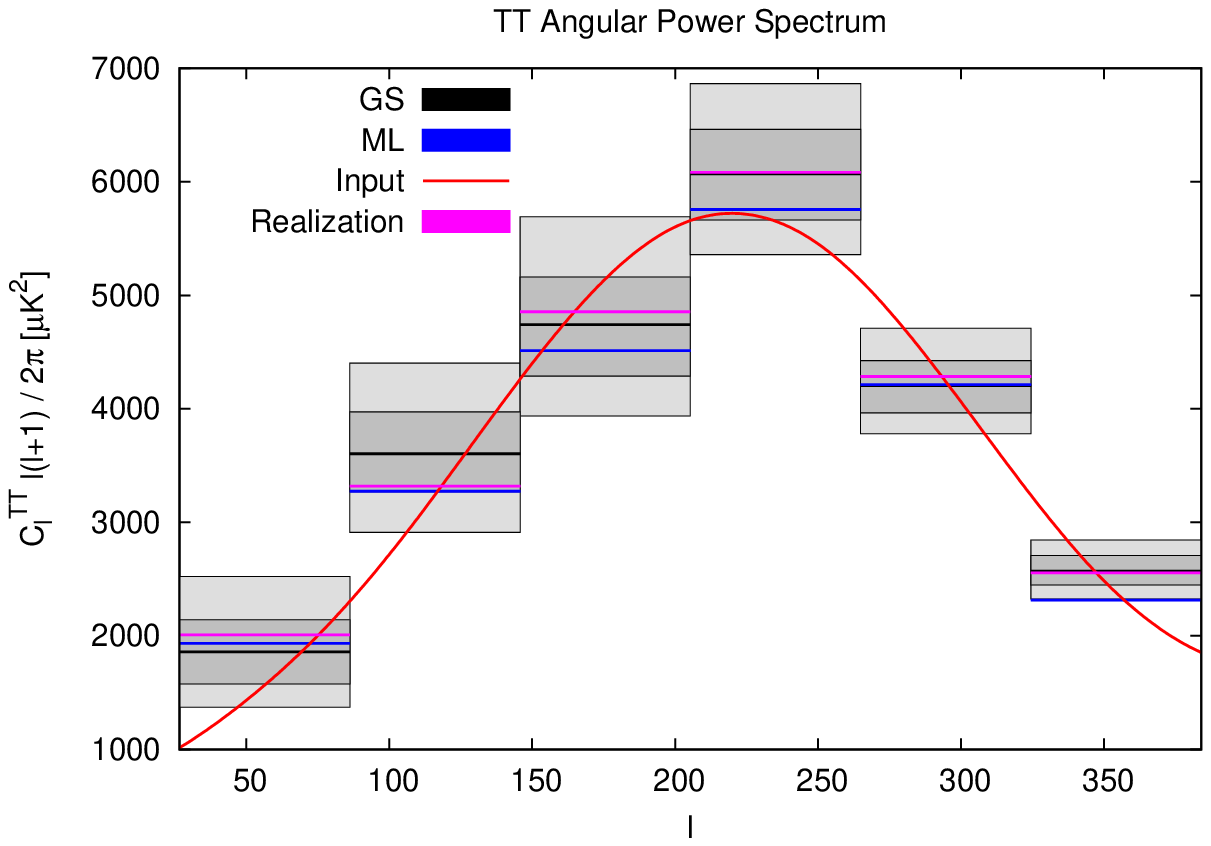} &
     \includegraphics[trim = 1mm 1mm 1mm 1mm, width=7.5cm]{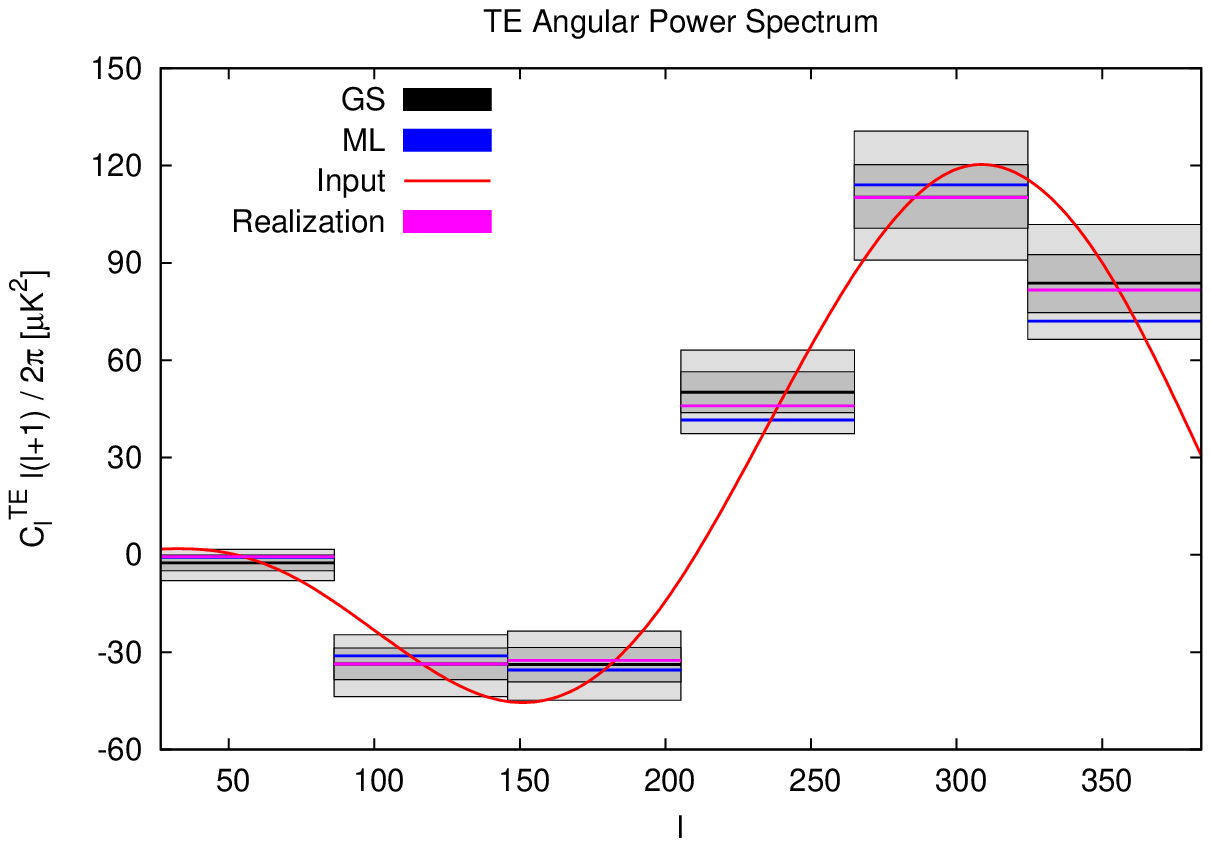} \\
     \includegraphics[trim = 1mm 1mm 1mm 1mm, width=7.5cm]{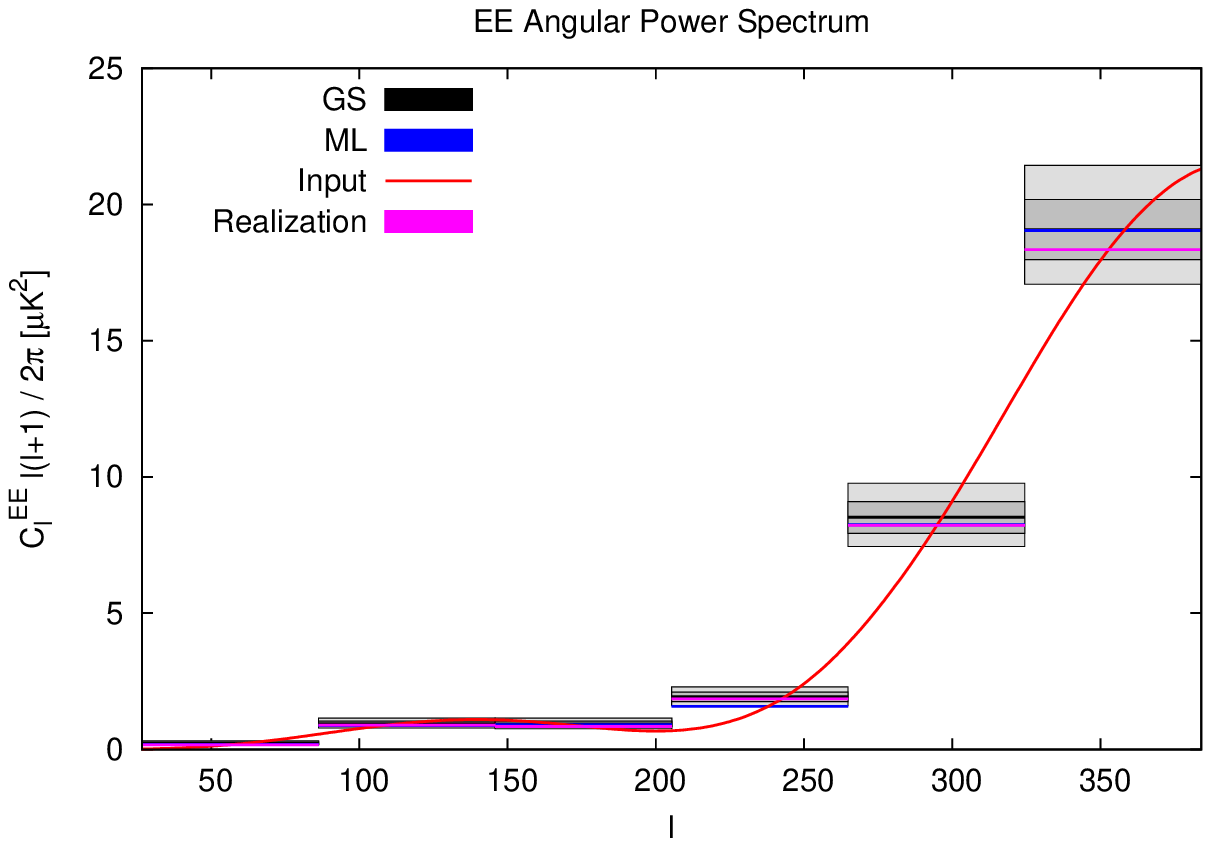} &
     \includegraphics[trim = 1mm 1mm 1mm 1mm, width=7.5cm]{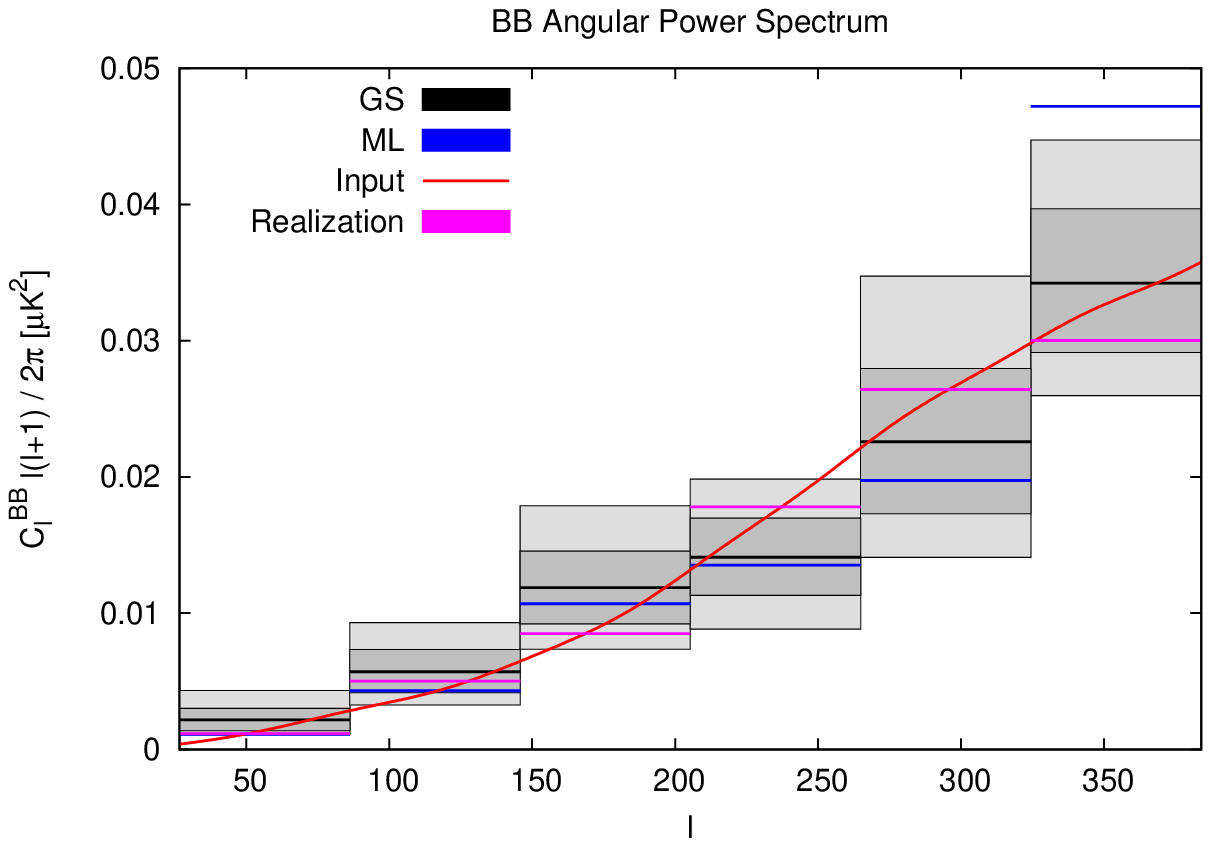} \\
     \includegraphics[trim = 1mm 1mm 1mm 1mm, width=7.5cm]{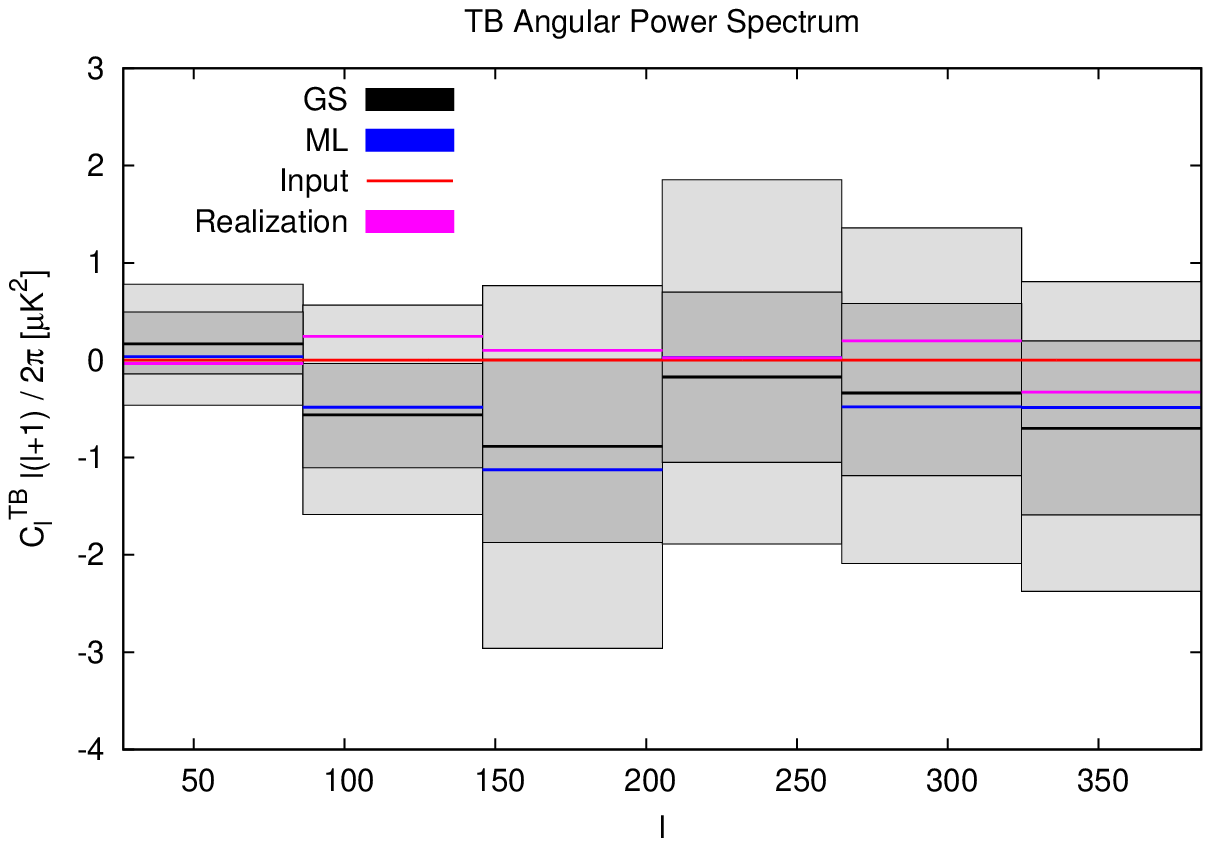} &
     \includegraphics[trim = 1mm 1mm 1mm 1mm, width=7.5cm]{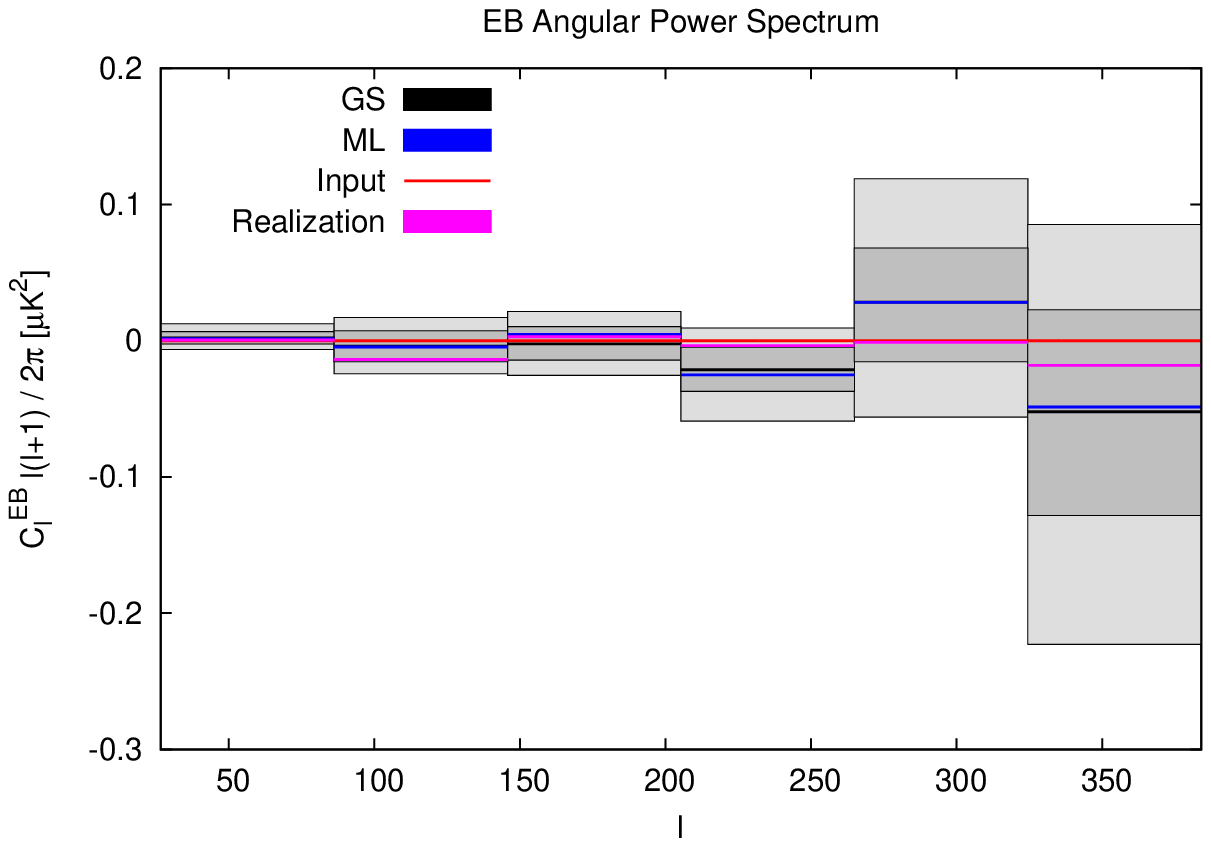} \\
     \end{array}$
       \caption{Mean posterior power spectra obtained by Gibbs Sampling (GS) for each $\ell$-bin are shown in black. The power spectra estimations obtained by Maximum Likelihood (ML) method are shown in blue. Dark and light grey indicate $1\sigma$ and $2\sigma$ uncertainties for Gibbs sampling results, respectively. The binned power spectra of the signal realization are shown in pink. Red lines are the input CMB power spectra obtained by CAMB for a tensor-to-scalar ratio of $r = 0.01$.}
  \end{center}
\end{figure*}

The mean posterior power spectra, together with the associated uncertainties at each $\ell$-bin, obtained by the methods of Gibbs Sampling (GS) and Maximum Likelihood (ML) for the ideal linear experiment, are shown in Figure 2. The input power spectra, which are used to construct the signal realization, and the spectra of the signal realization are also shown in Figure 2. Almost all of our estimates fall within $2\sigma$ of the expected value.

\subsection{Effect of Errors}

In order to estimate $\alpha$ we ran 30 realizations of each systematic error simulation for both linear and circular experiments. To keep the value of $\alpha_r$ less than $10\%$ tolerance limit at $r=0.01$, we set the rms values of the parameters for gain errors to $|g_{rms}| = 0.1$, for coupling errors to $|\epsilon_{rms}| = 5 \times 10^{-4}$, for pointing errors to $\delta_{rms} \approx 0.7^\circ$, for beam shape errors to $\zeta_{rms} \approx 0.7^\circ$, and for cross-polarization errors to $\mu_{rms} = 5 \times 10^{-4}$.
\begin{figure*} \label{fig:beamerr}
  \begin{center}$
    \leavevmode
    \begin{array}{c@{\hspace{.1cm}}c@{\hspace{.1cm}}c}
     \includegraphics[trim = 3mm .1mm 2mm 2mm, clip=true, width=5.8cm]{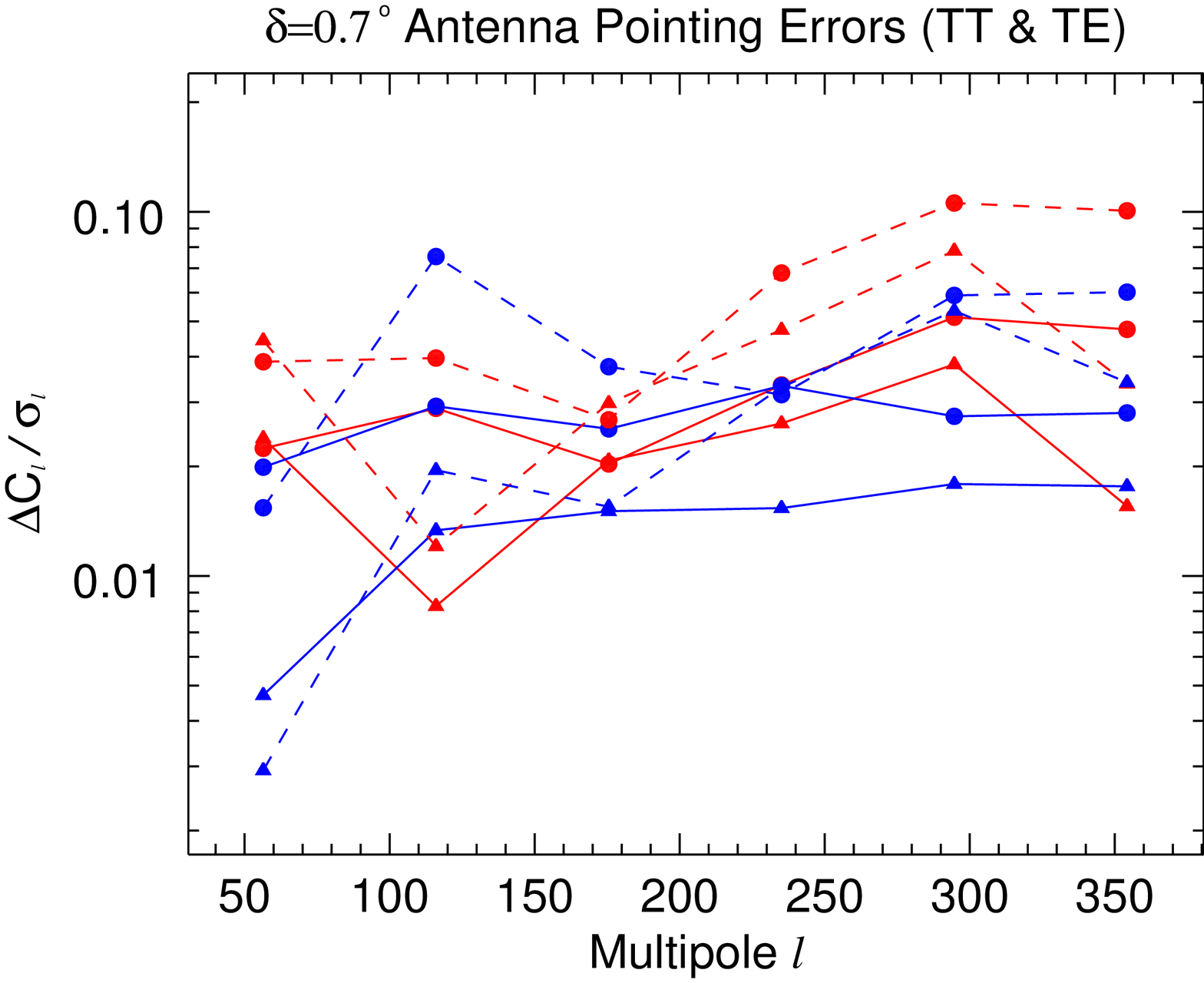} &
     \includegraphics[trim = 2mm .1mm 2mm 2mm, clip=true, width=5.8cm]{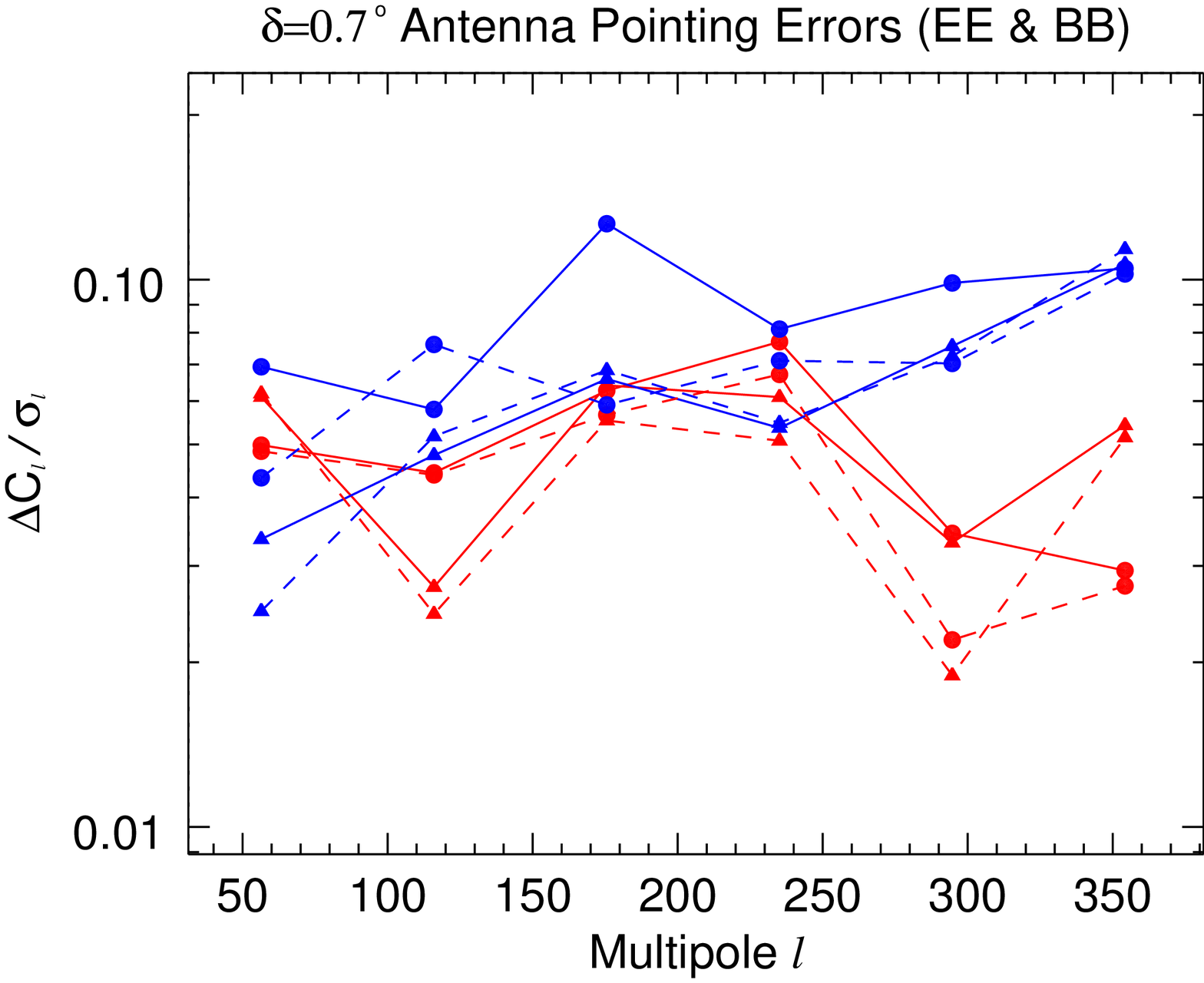} &
     \includegraphics[trim = 2mm .1mm 2mm 2mm, clip=true, width=5.8cm]{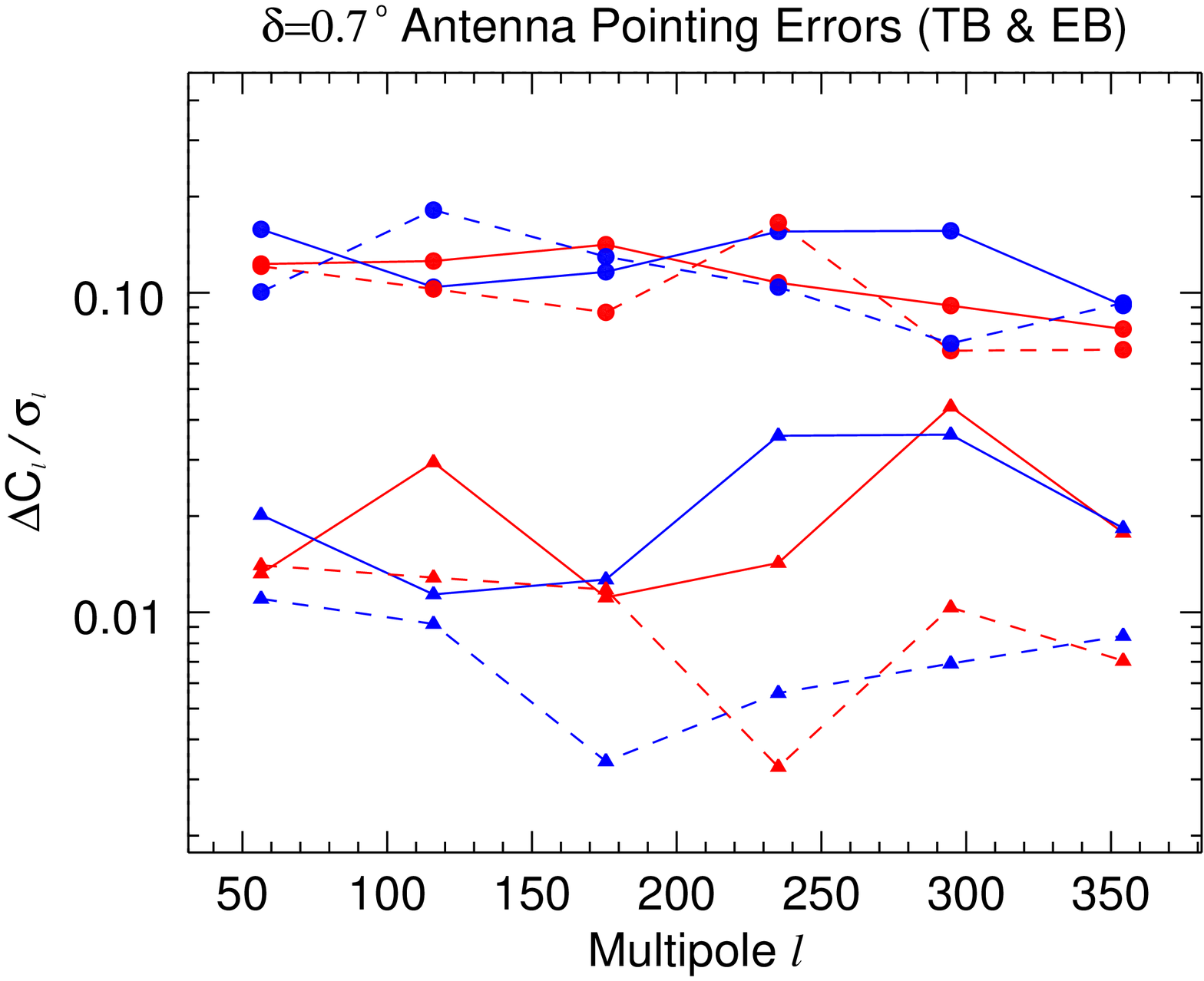} \\
     \includegraphics[trim = 3mm .1mm 2mm 2mm, clip=true, width=5.8cm]{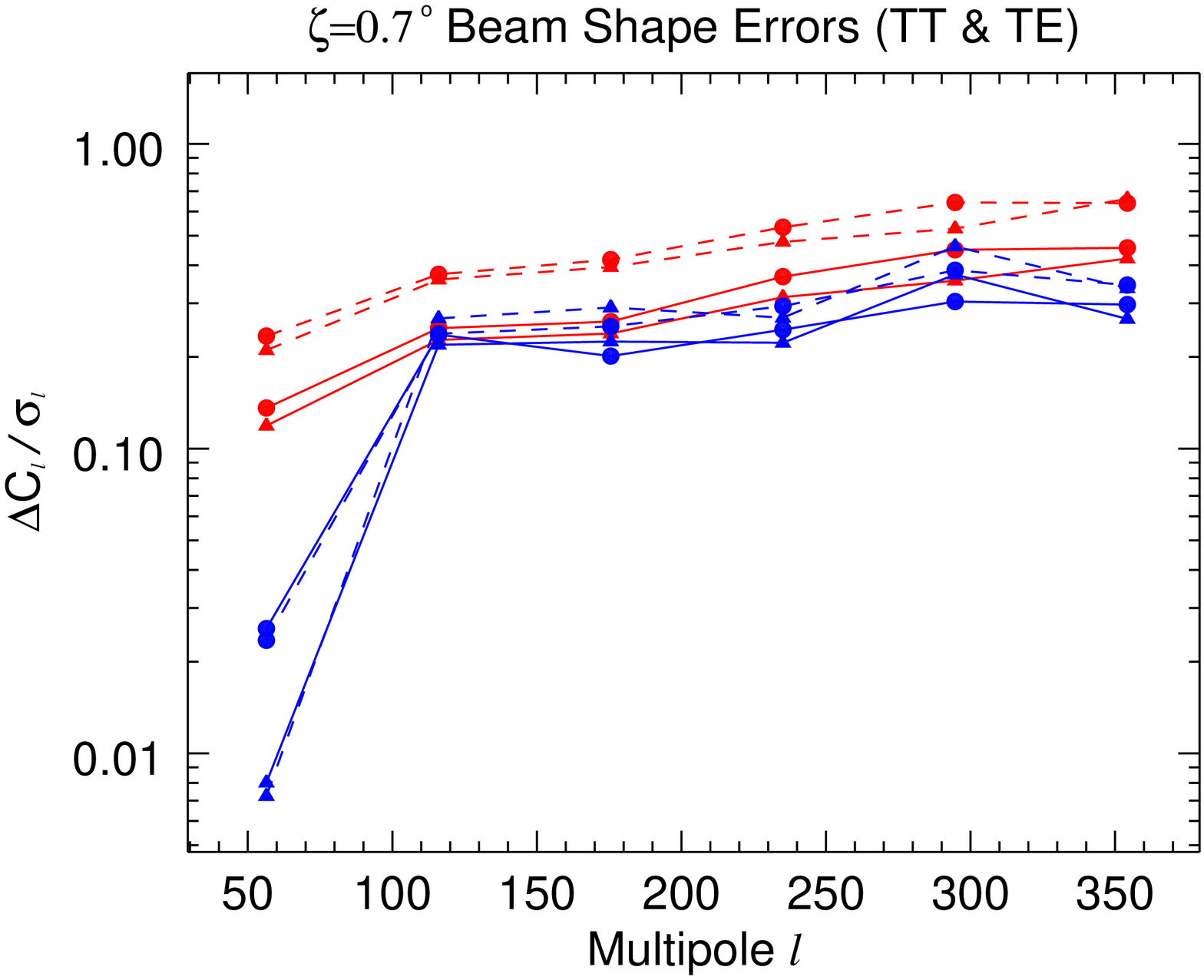} &
     \includegraphics[trim = 2mm .1mm 2mm 2mm, clip=true, width=5.8cm]{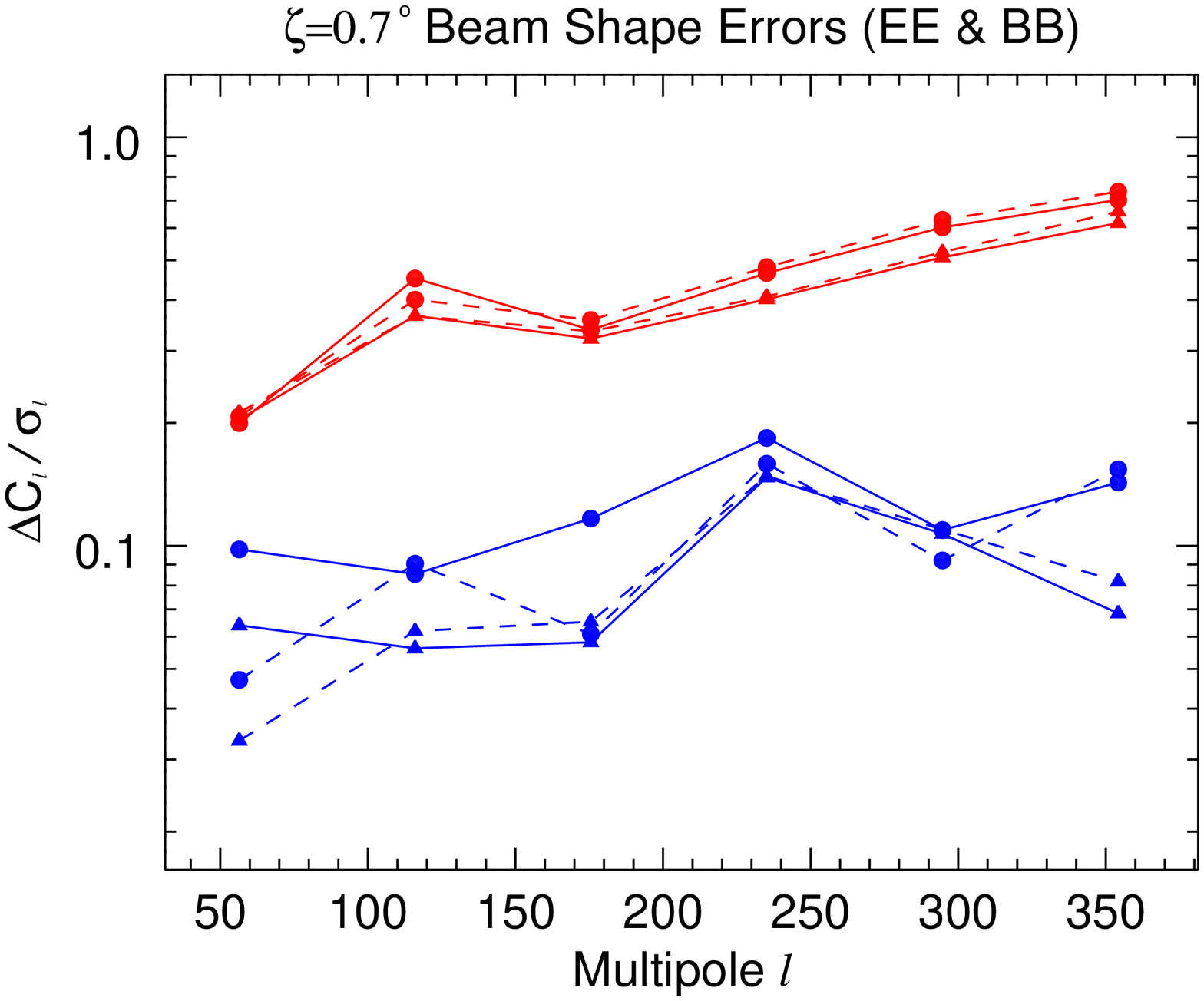} &
     \includegraphics[trim = 2mm .1mm 2mm 2mm, clip=true, width=5.8cm]{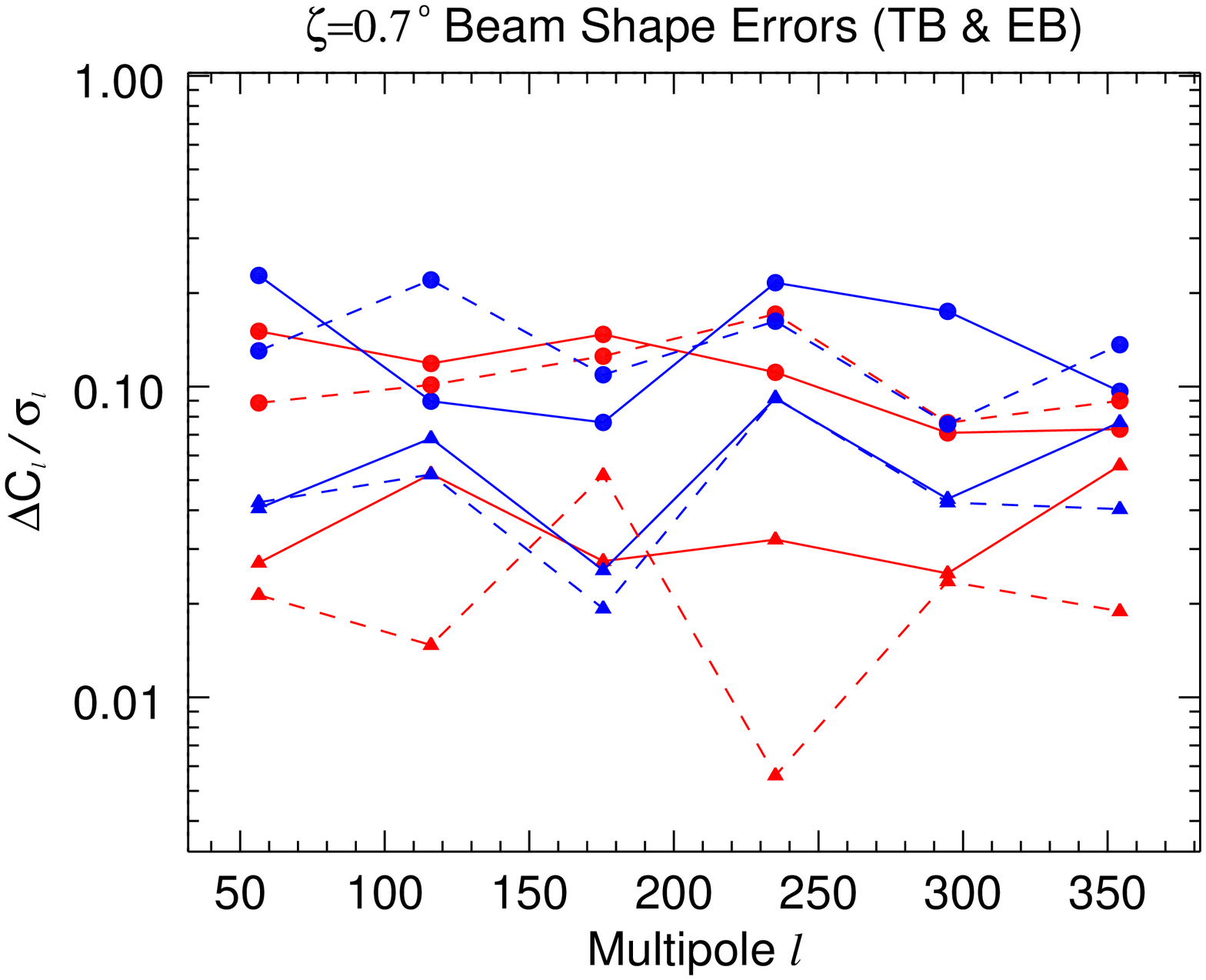} \\
     \includegraphics[trim = 3mm .1mm 2mm 2mm, clip=true, width=5.8cm]{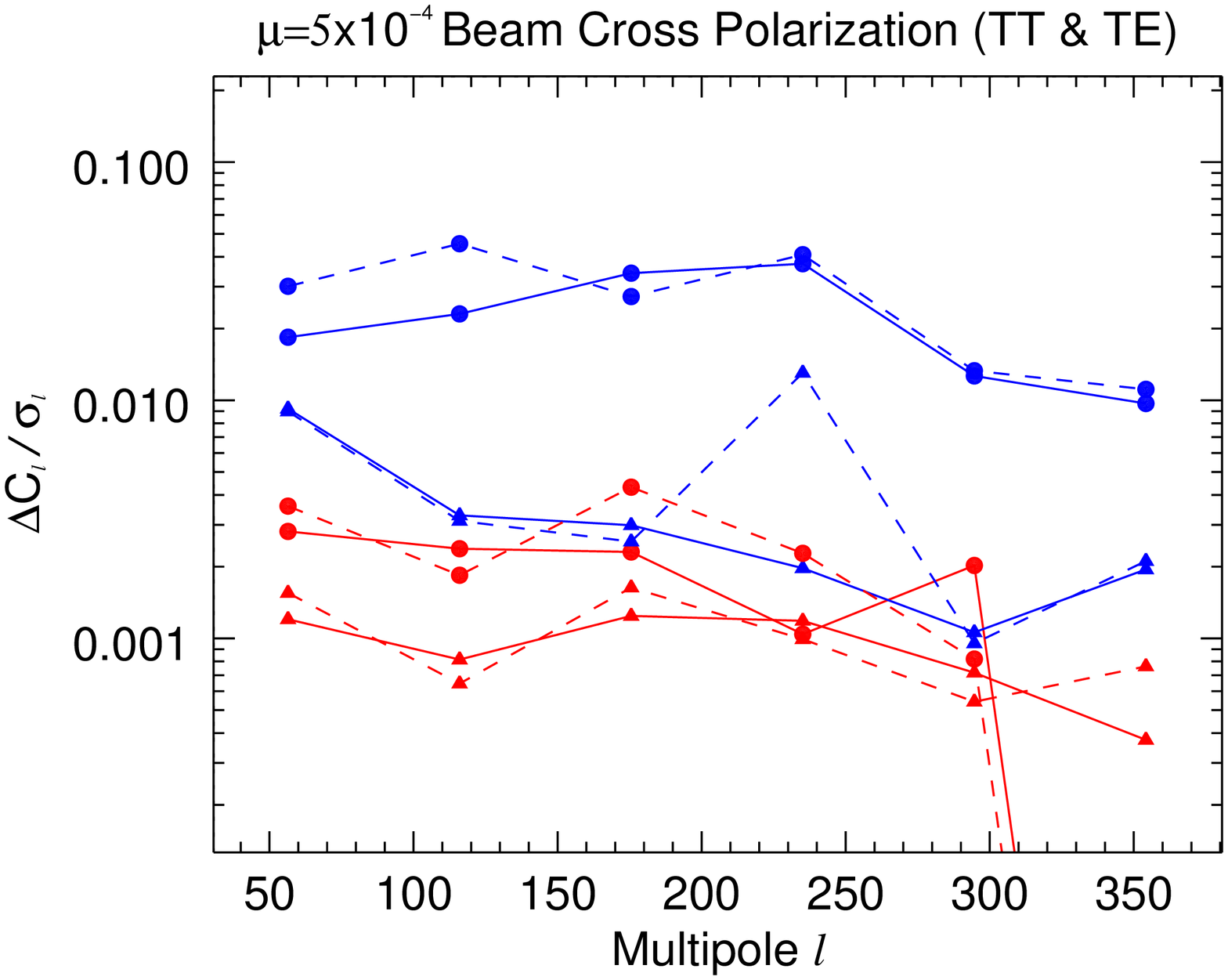} &
     \includegraphics[trim = 2mm .1mm 2mm 2mm, clip=true, width=5.8cm]{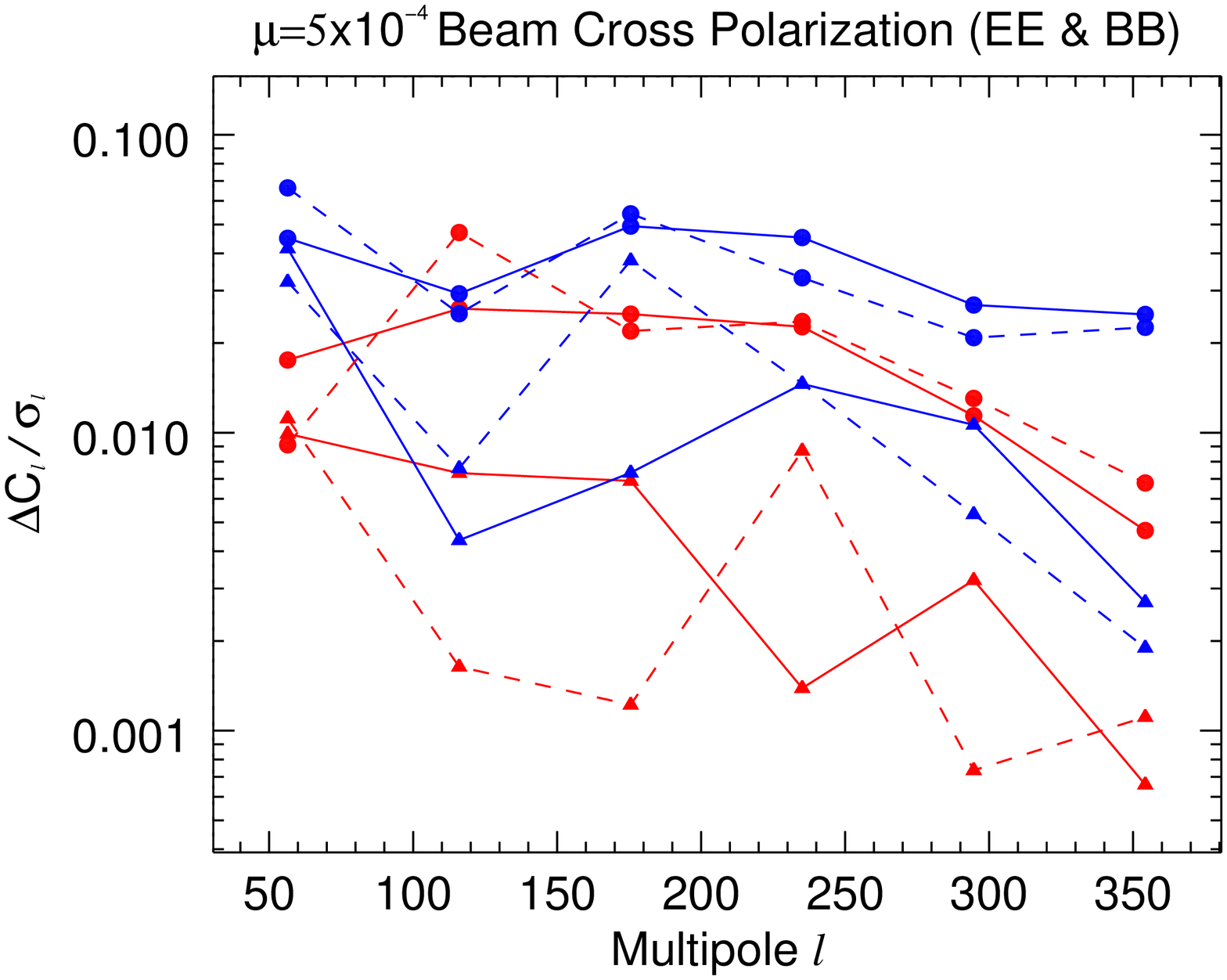} &
     \includegraphics[trim = 2mm .1mm 2mm 2mm, clip=true, width=5.8cm]{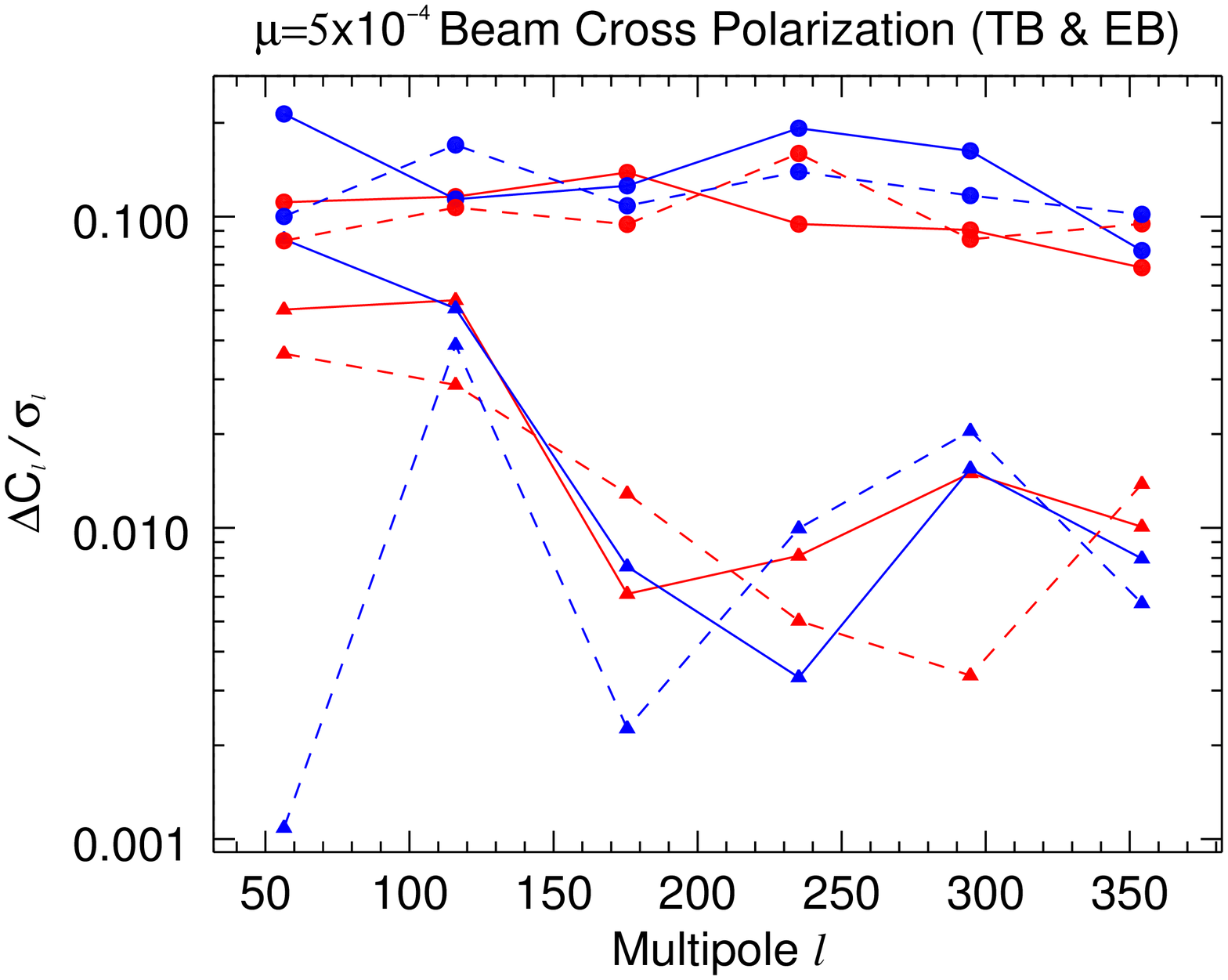} \\
     \end{array}$
       \caption{Beam errors. The values of $\alpha^{XY}$, averaged over 30 simulations, obtained by both maximum likelihood (ML) method (triangles) and the method of Gibbs sampling (GS) (solid dots) are shown. The three rows indicate, from top to bottom, pointing errors with $\delta_{rms} \approx 0.7^\circ$, beam shape errors with $\zeta_{rms} \approx 0.7^\circ$, and beam cross-polarization with $\mu_{rms} = 5 \times 10^{-4}$. Left panel shows $\alpha^{TT}$ (red) and $\alpha^{TE}$ (blue). Middle panel shows $\alpha^{EE}$ (red) and $\alpha^{BB}$ (blue). Right panel shows $\alpha^{TB}$ (red) and $\alpha^{EB}$ (blue). Linear and circular experiments are shown by solid and dashed lines, respectively.}
  \end{center}
\end{figure*}

\begin{figure*} \label{fig:insterr}
  \begin{center}$
    \leavevmode
    \begin{array}{c@{\hspace{.1cm}}c@{\hspace{.1cm}}c}
     \includegraphics[trim = 2mm .1mm 2mm 2mm, clip=true, width=5.8cm]{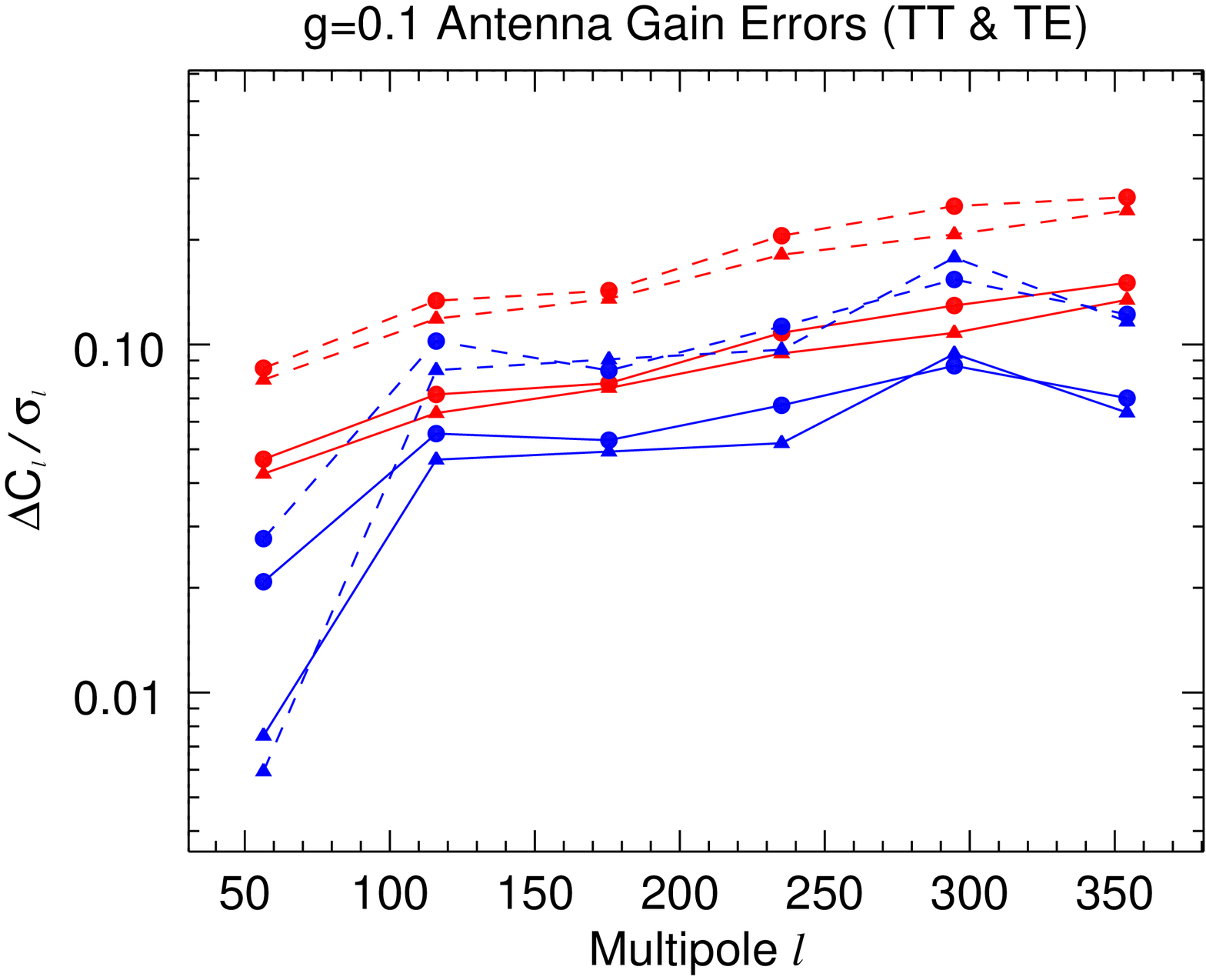} &
     \includegraphics[trim = 2mm .1mm 2mm 2mm, clip=true, width=5.8cm]{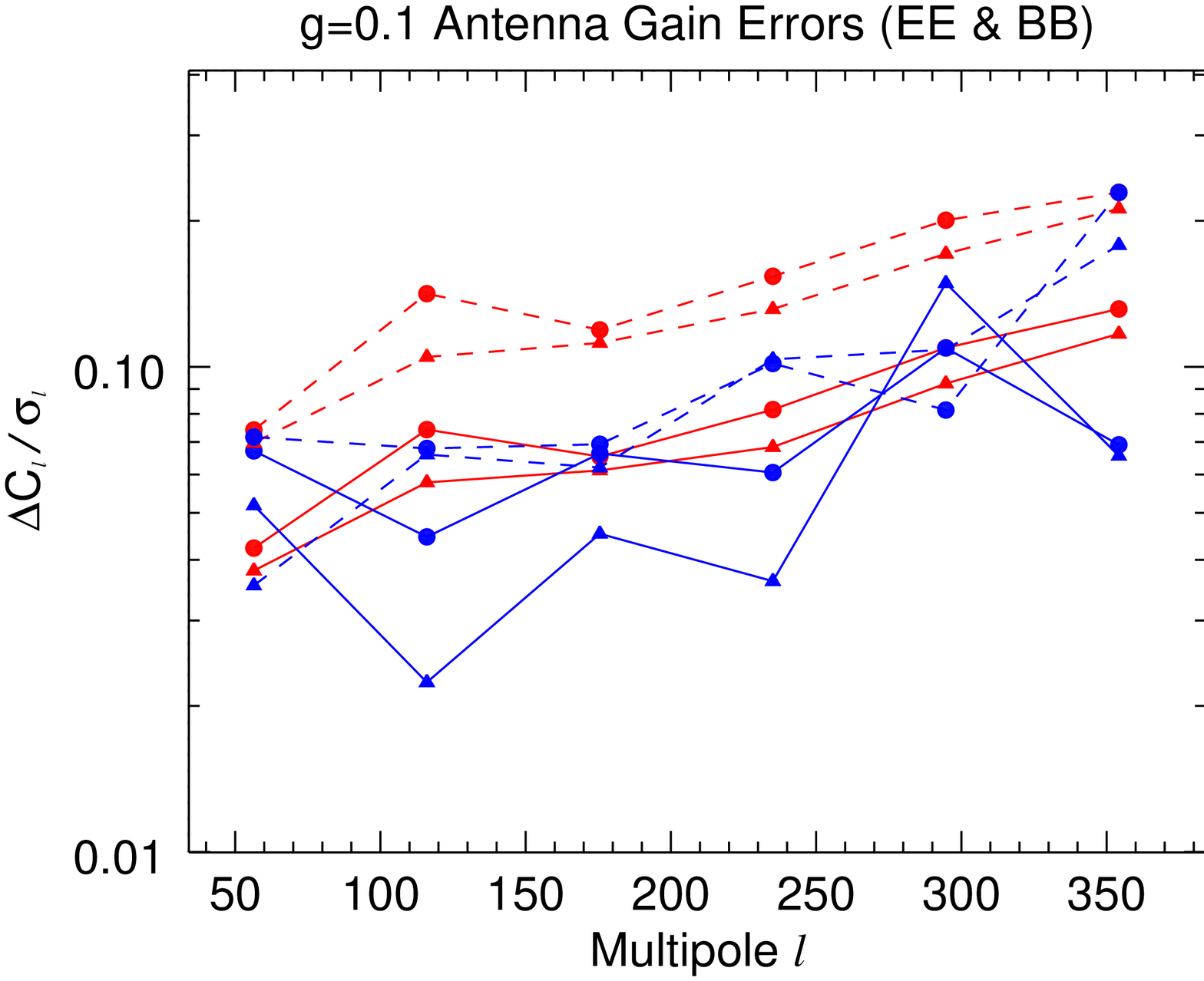} &
     \includegraphics[trim = 2mm .1mm 2mm 2mm, clip=true, width=5.8cm]{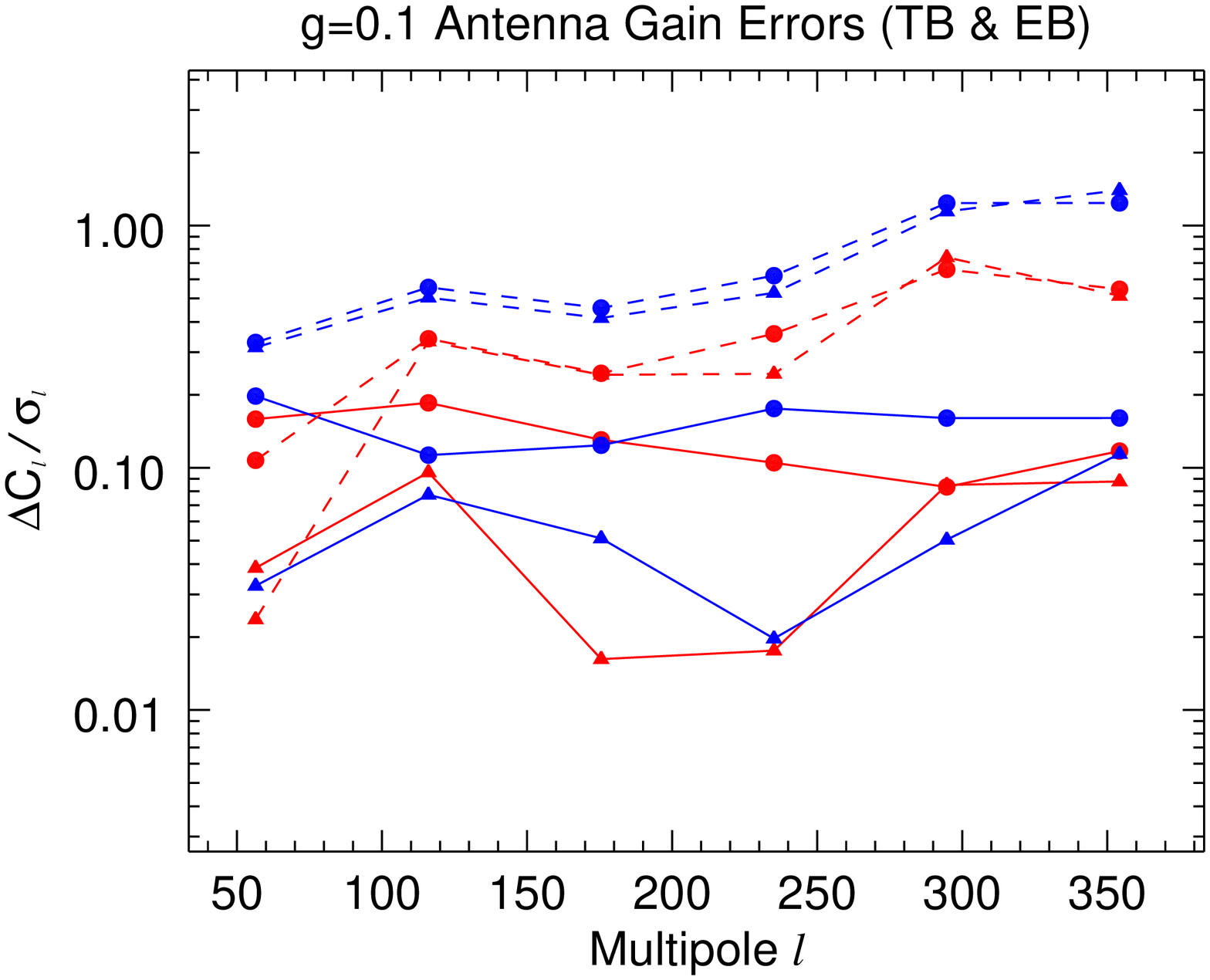} \\
     \includegraphics[trim = 2mm .1mm 2mm 2mm, clip=true, width=5.8cm]{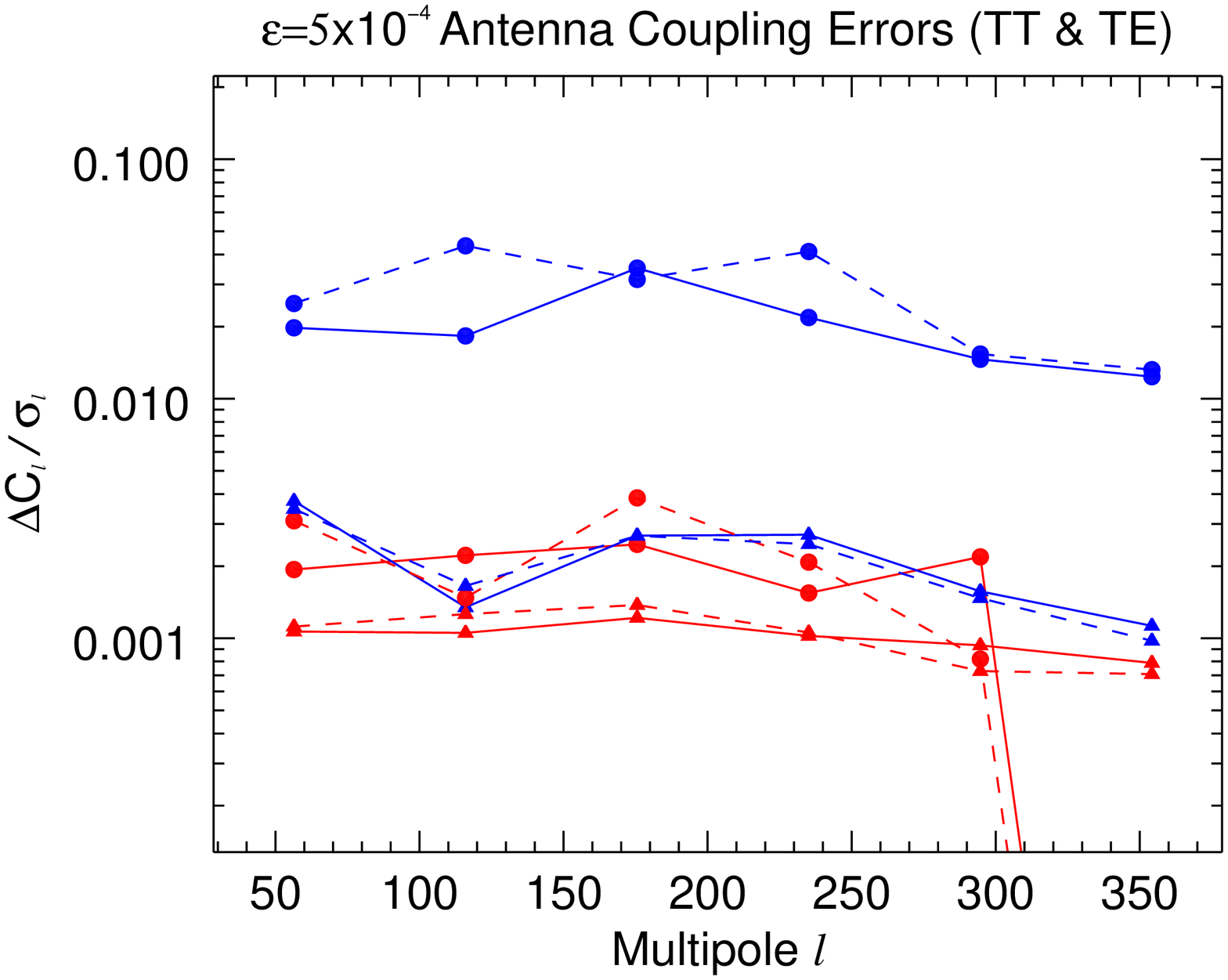} &
     \includegraphics[trim = 2mm .1mm 2mm 2mm, clip=true, width=5.8cm]{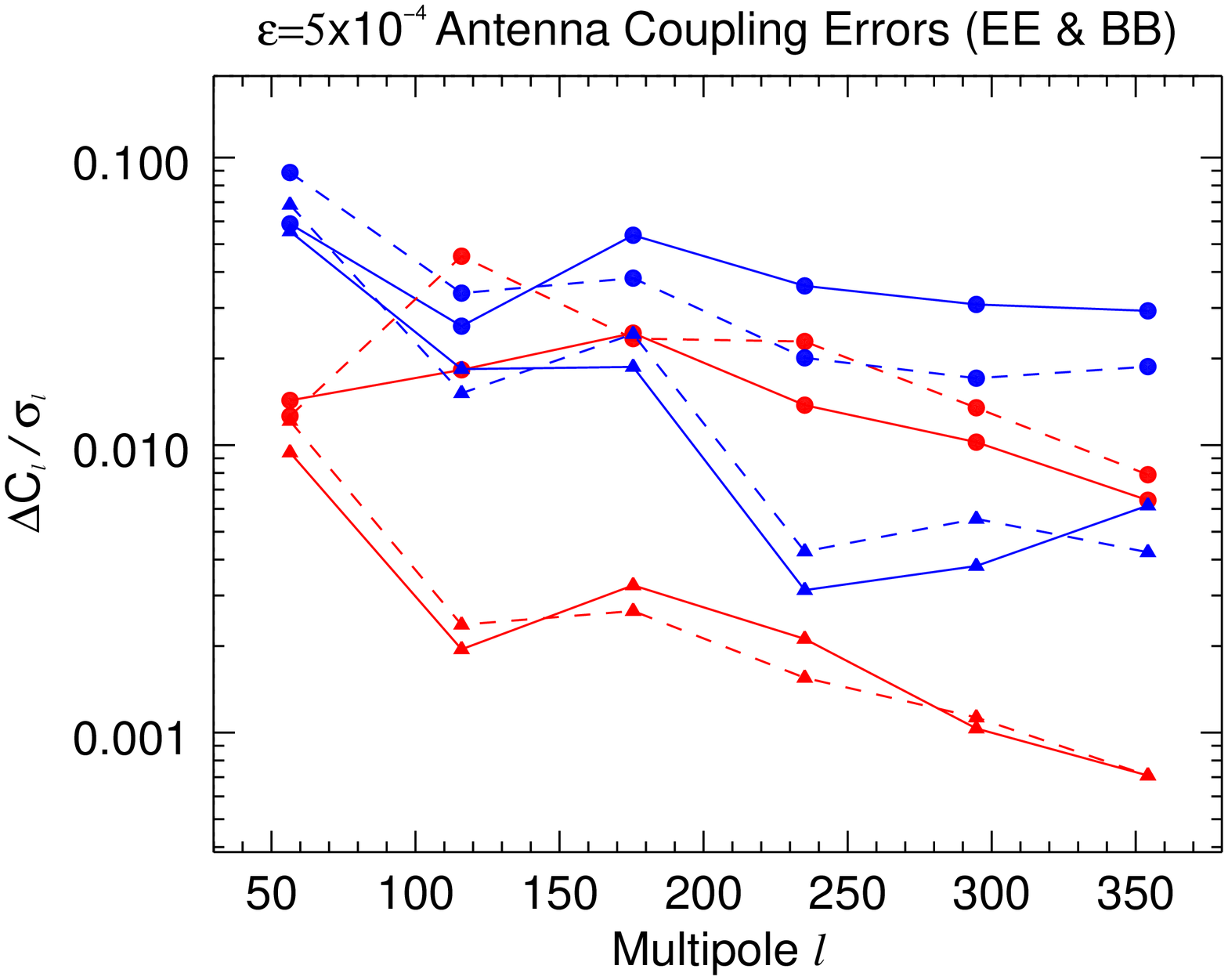} &
     \includegraphics[trim = 2mm .1mm 2mm 2mm, clip=true, width=5.8cm]{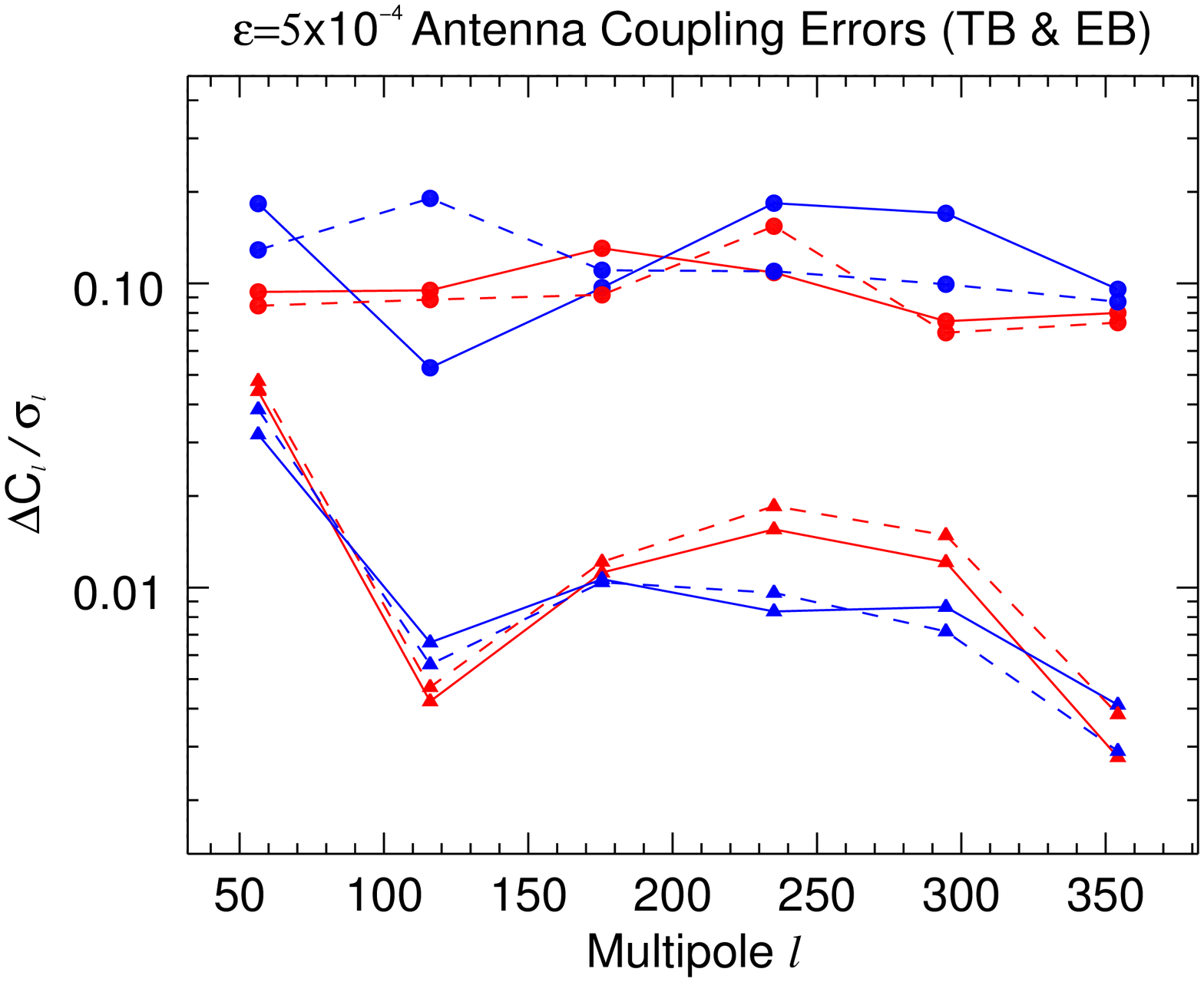} \\
     \includegraphics[trim = 2mm .1mm 2mm 2mm, clip=true, width=5.8cm]{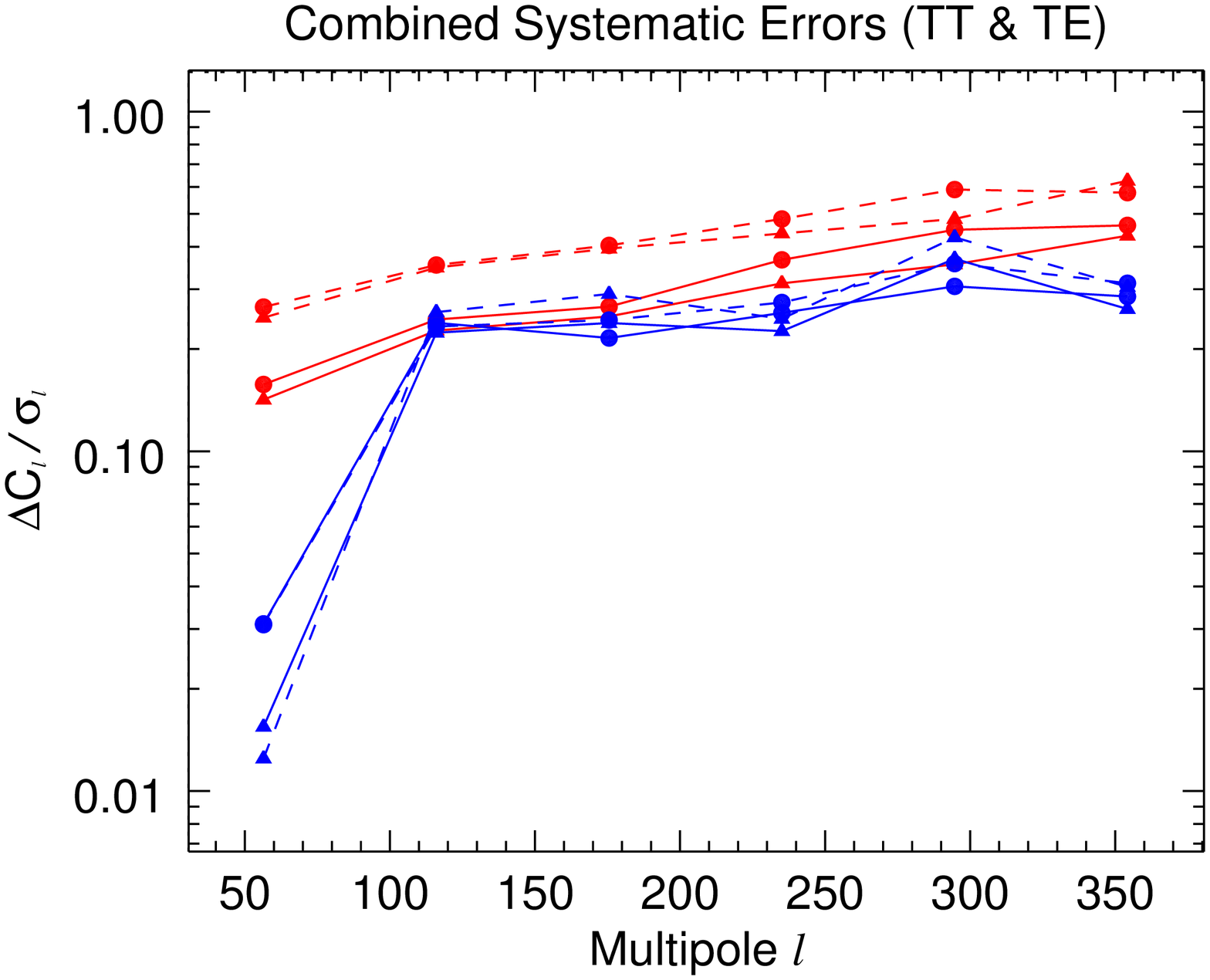} &
     \includegraphics[trim = 2mm .1mm 2mm 2mm, clip=true, width=5.8cm]{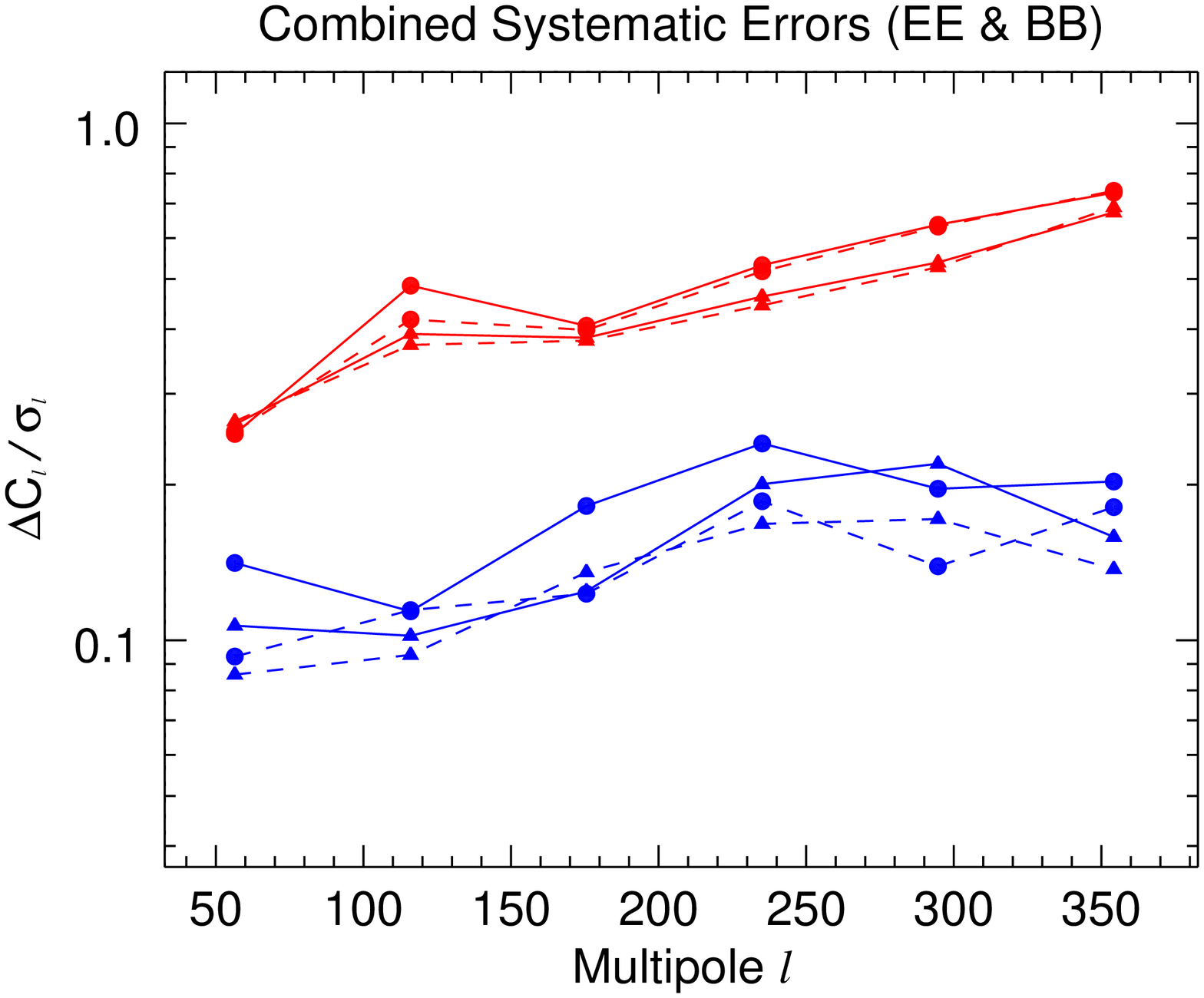} &
     \includegraphics[trim = 2mm .1mm 2mm 2mm, clip=true, width=5.8cm]{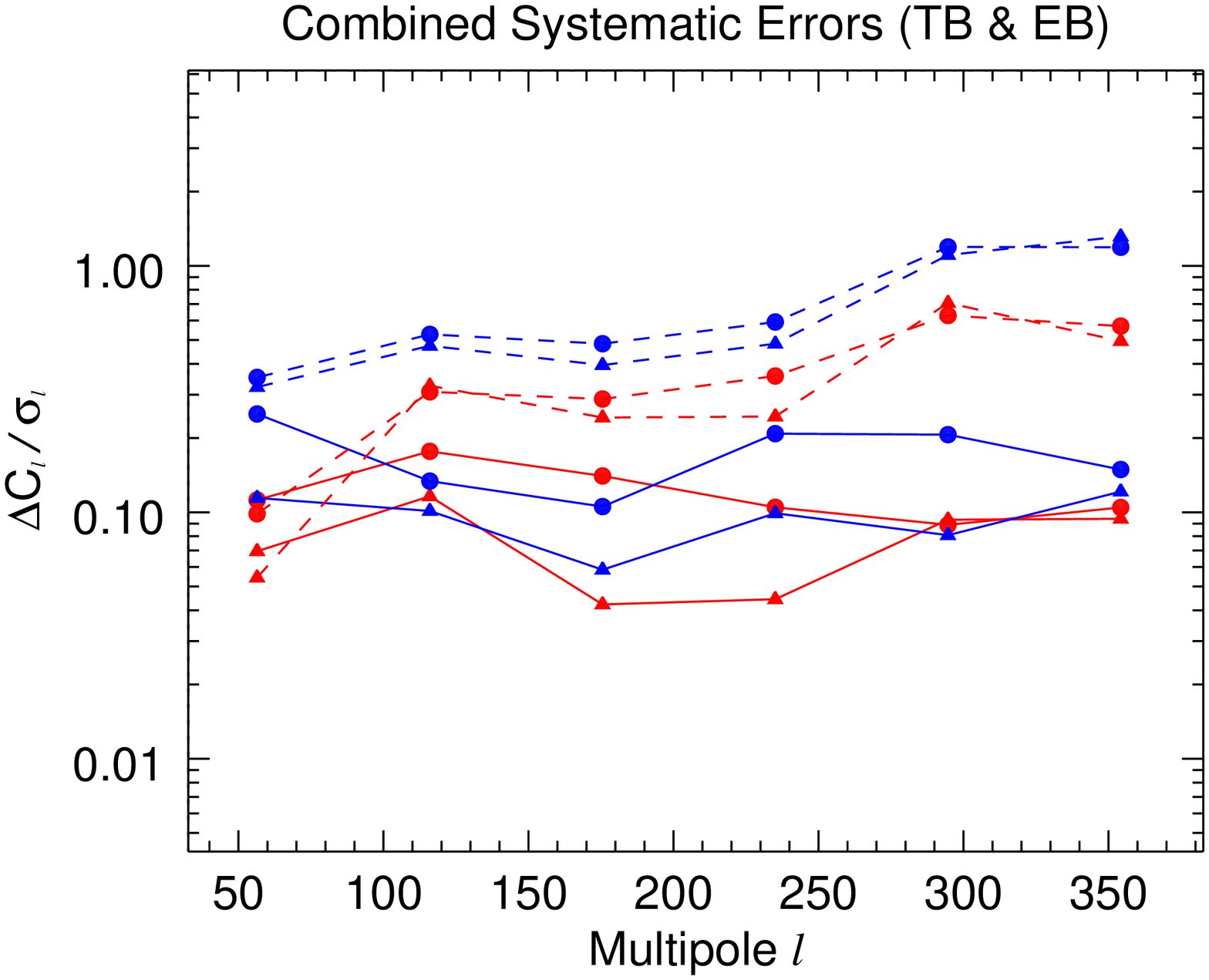} \\
     \end{array}$
       \caption{Instrumental and combined systematic errors. The values of $\alpha^{XY}$, averaged over 30 simulations, obtained by both maximum likelihood (ML) method (triangles) and the method of Gibbs sampling (GS) (solid dots) are shown. Top row:  antenna gain with $|g_{rms}| = 0.1$. Middle row: antenna couplings with $|\epsilon_{rms}| = 5\times 10^{-4}$. Bottom row: combined effect of beam and instrumental systematic errors. Left panel shows $\alpha^{TT}$ (red) and $\alpha^{TE}$ (blue). Middle panel shows $\alpha^{EE}$ (red) and $\alpha^{BB}$ (blue). Right panel shows $\alpha^{TB}$ (red) and $\alpha^{EB}$ (blue). Linear and circular experiments are shown by solid and dashed lines, respectively.}
  \end{center}
\end{figure*}

%Since the $TB$ and $EB$ spectra from ideal experiments roughly vanish, we show $\delta C^{TB}$ and $\delta C^{EB}$ instead of $\alpha^{TB}$ and $\alpha^{EB}$.

Figure 3 shows the mean values of $\alpha^{XY}$ for beam errors, averaged over 30 realizations. The results from ML and GS methods are in good agreement for both linear and circular experiments. In all three cases $\alpha^{BB} \sim 0.1$ at low $\ell$, as expected. Although the cross polarization has a much smaller error parameter, its effect on the power spectra is comparable to the pointing and shape errors. The reason for this is the leakage from $TT$ power into $BB$ power that is caused by the off-diagonal elements of the beam pattern, whereas the source of $\alpha^{BB}$ for pointing and shape errors is the $EE \to BB$ leakage (Bunn 2007).  

The mean values of $\alpha^{XY}$ for instrumental errors are shown in Figure 4. For gain and coupling errors, $\alpha^{XY}$ is roughly at the $10\%$ level. The main contribution for the $\alpha^{BB}$ comes from the leakage from $EE$ power into $BB$ power for gain errors. As in the case of cross-polarization errors, despite having a much smaller parameter than gain, $\alpha^{BB} \sim 0.1$ at low $\ell$ for antenna coupling errors because of $TT \to BB$ leakage. 

We simulated the systematics by turning on one error at a time. However, in a realistic experiment, all systematic errors act together simultaneously, causing a larger effect on the spectra. In order to see this combined effect we ran 30 realizations with all the systematic errors discussed in previous sections turned on at once. The results are also shown in Figure 4. As expected, the combined effect is almost twice as large as the individual cases.

\subsection{Comparison to Analytical Estimations}

\begin{figure*} \label{fig:bar}

\centering
\mbox{
$\leavevmode
  \begin{array}{c@{\hspace{1cm}}c}
   \includegraphics[scale =0.28] {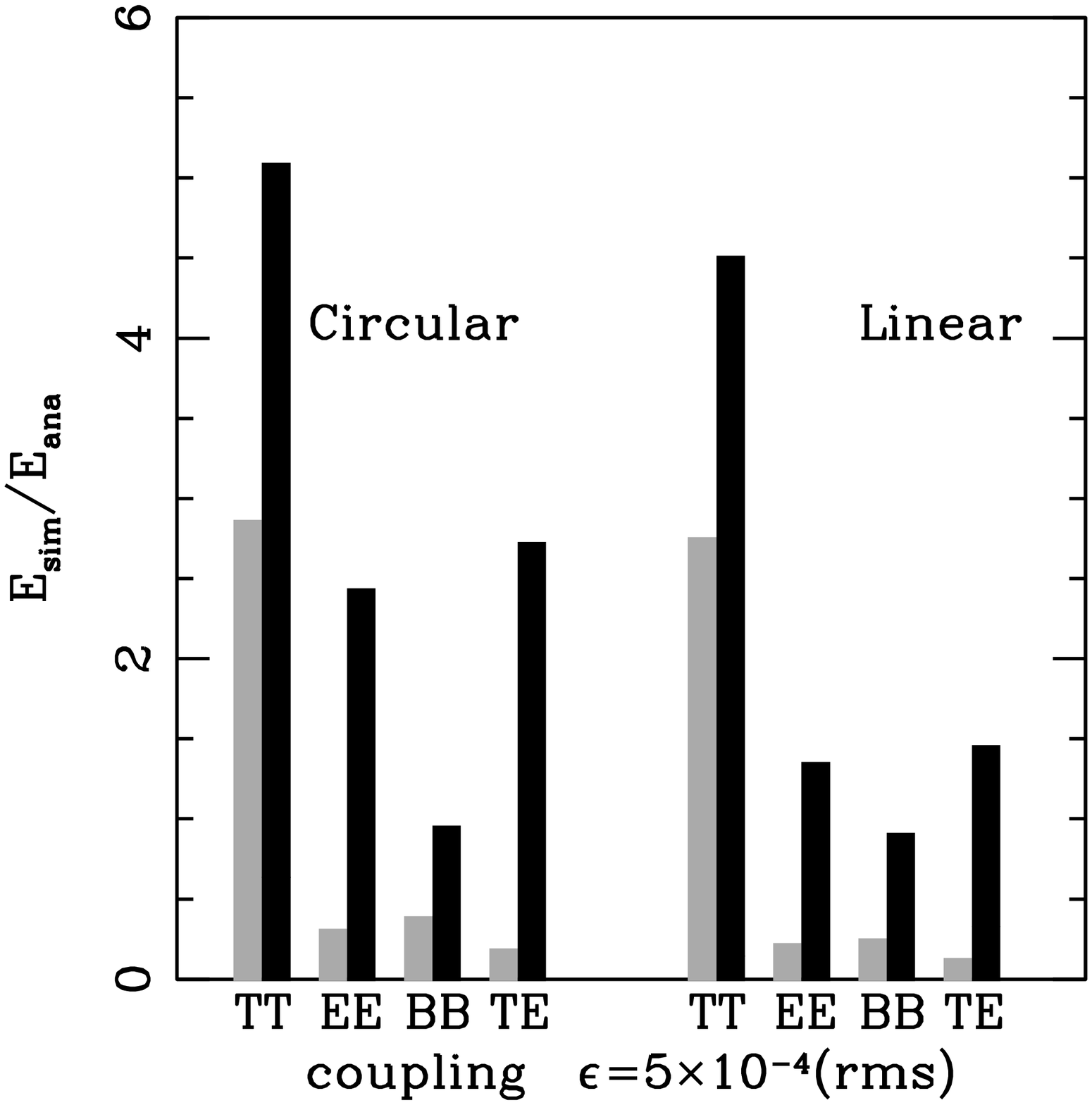} &
    \includegraphics[scale =0.28] {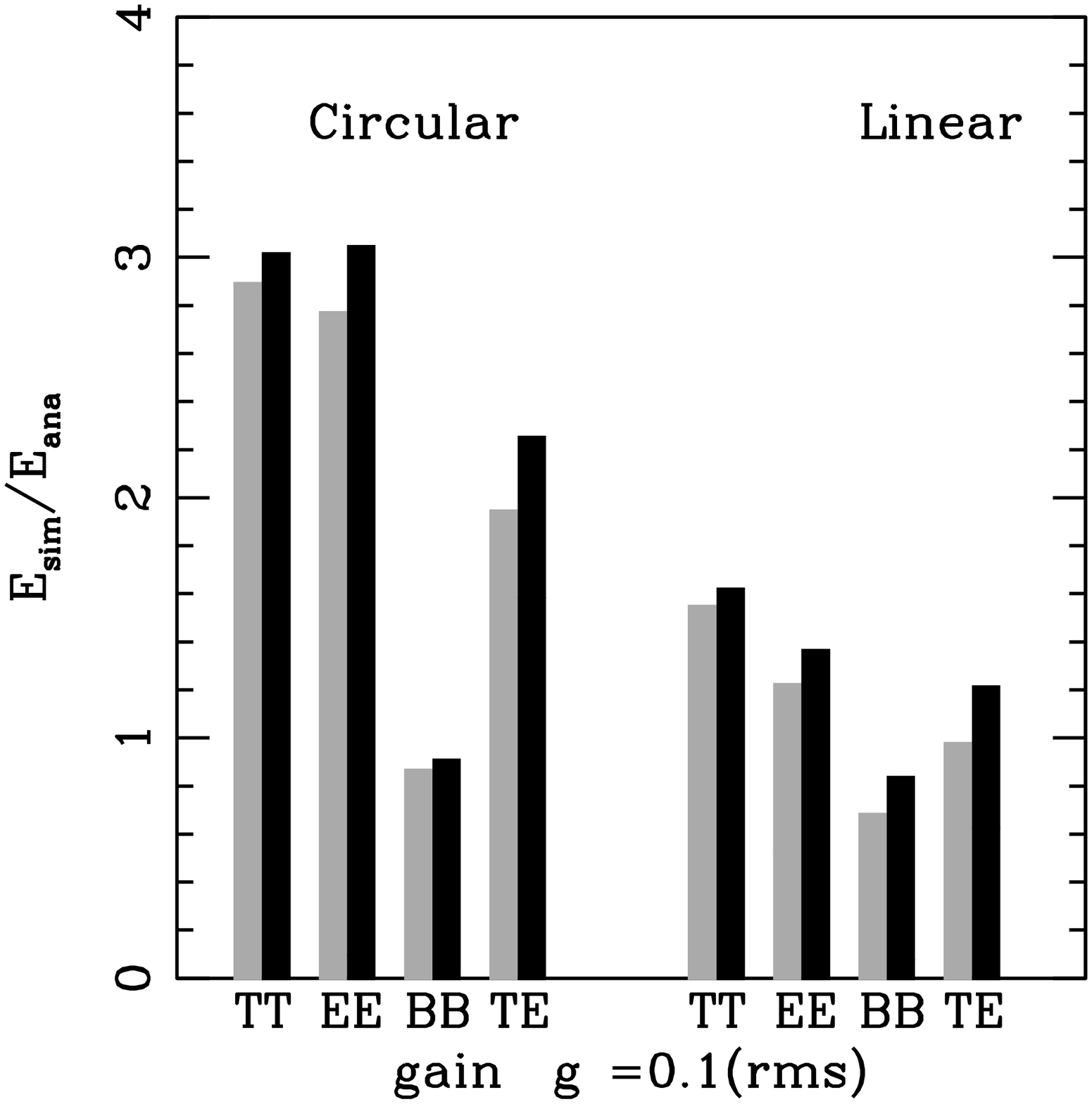}
   \end{array}$

}

\mbox{
$ \leavevmode
   \begin{array}{ccc}
   \includegraphics[scale =0.28] {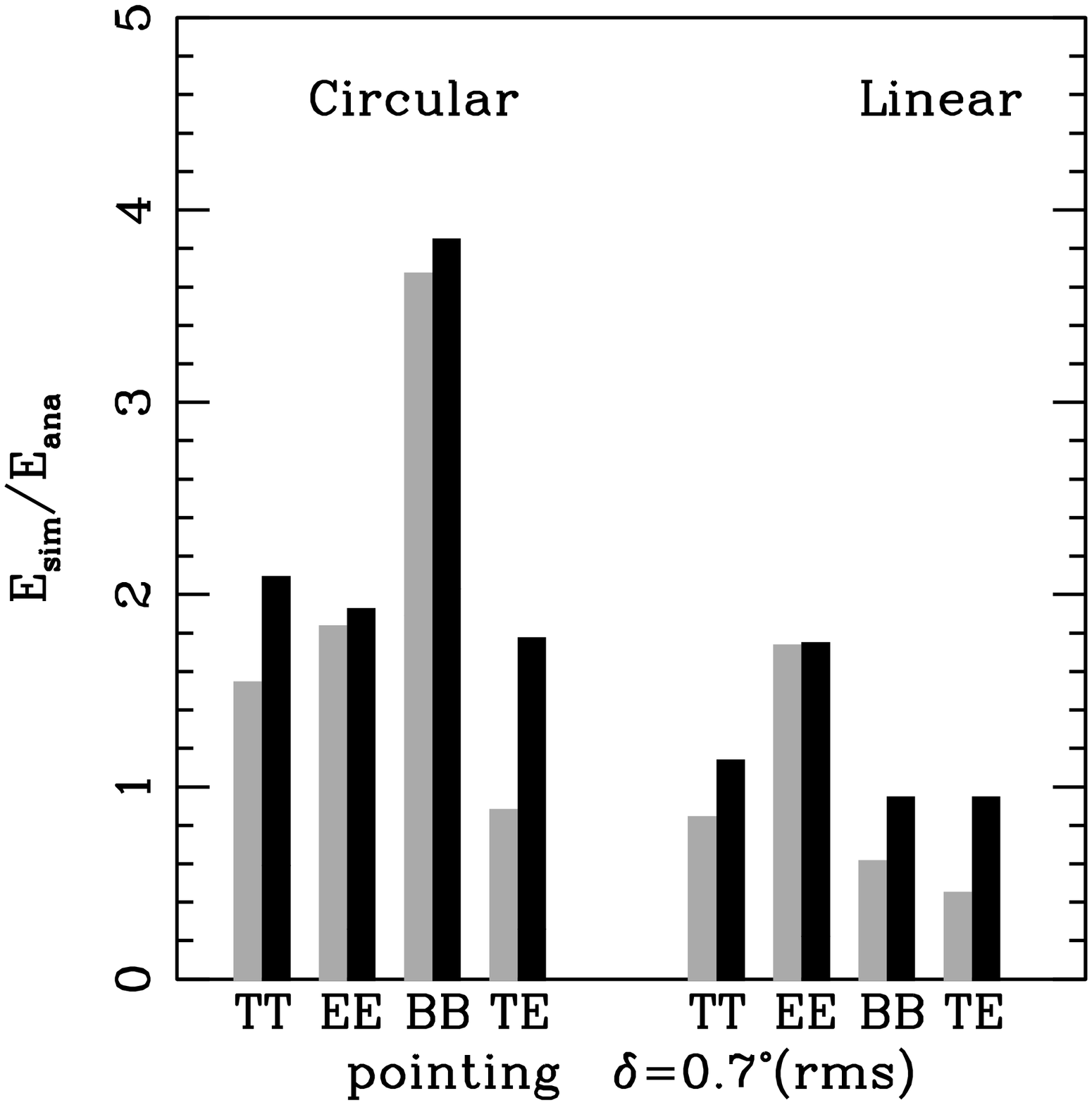} &
   \includegraphics[scale =0.28] {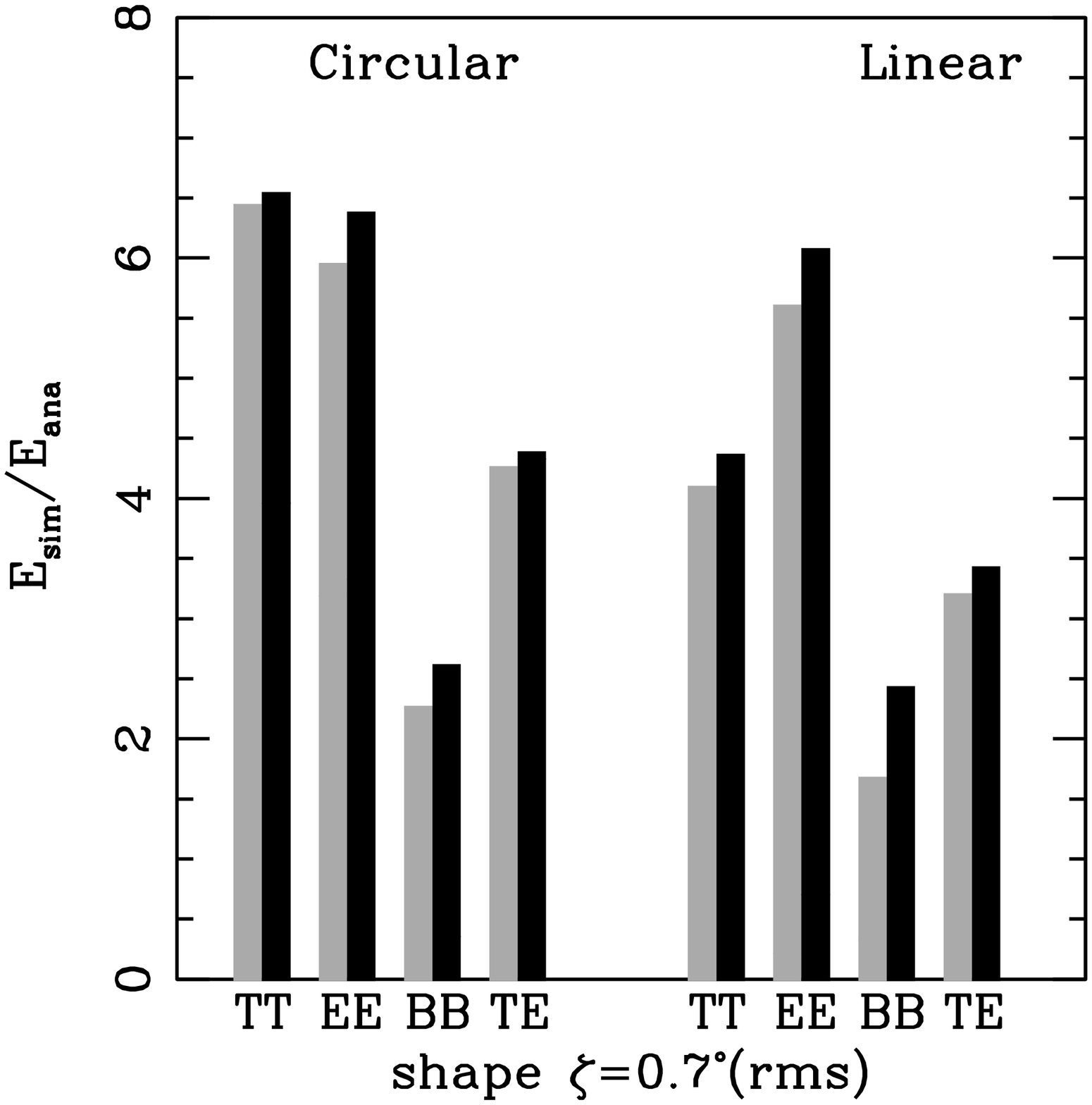} &
   \includegraphics[scale =0.28] {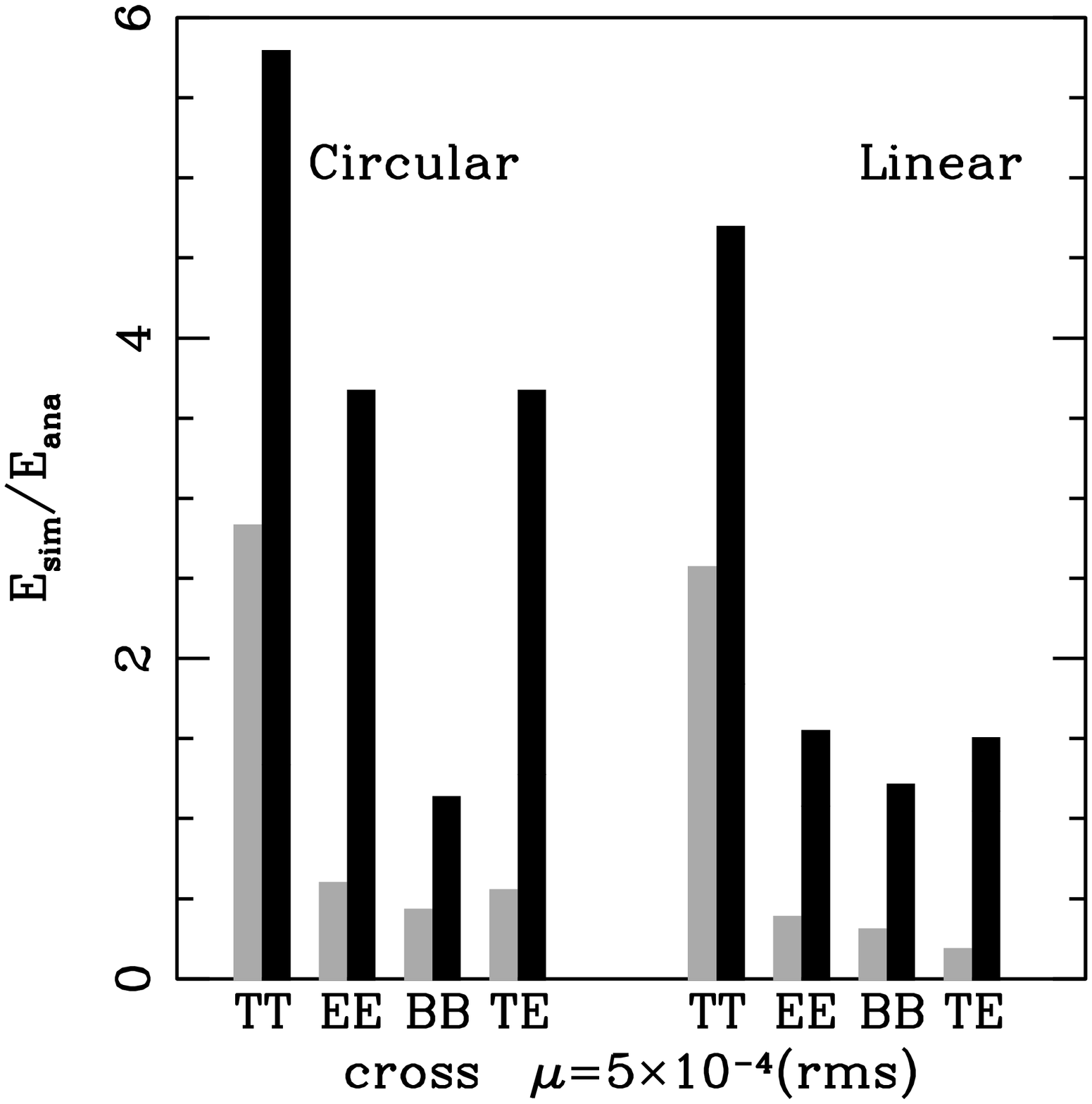}
   \end{array}$
}
\caption{The bin-averaged ratios of the simulated and analytical systematic errors. In each panel, the results from the simulations in both the linear and circular bases for $TT$, $EE$, $BB$, $TE$ are shown. The gray and black bars correspond to the maximum-likelihood and Gibbs-sampling methods in analysis of the simulated data, respectively.}
\end{figure*}

Analytical estimations for $\alpha^{XY}$ are obtained from the quadrature difference of Eq.~\ref{eq:delce}, normalized by the number of baselines. In general $\alpha^{XY}$ has a polynomial dependence on $\overline{s^2}$. For our interferometer configuration $\overline{s^2}$ is roughly $\overline{s^2} \sim 262.7 / \ell^2$. The explicit forms of the unnormalized estimations are given in the Appendix.

In general, our simulated results are larger than the estimations in all $\ell$-bins. This is expected because our analytical estimations are only first order approximations where it is also assumed that the errors associated with baselines are uncorrelated, making them lower bounds for the estimations. In reality, there is a correlation between errors associated with baselines having common antennas, a fact that is captured by our simulations. Upper bounds for the estimations can be found by unrealistically assuming full correlation of errors between baselines, where each baseline has the same error. For our interferometer design, this corresponds to roughly 65 times larger values. We expect our results to fall between uncorrelated and fully correlated estimations. In order to compare our results with the analytical ones, we consider the rms values of $\alpha^{XY}$ averaging over the $\ell$-bins. Figure 5 shows the ratios of $\alpha^{XY}_{rms}$ obtained by ML and GS methods to the estimated $\alpha^{XY}_{rms}$. In most cases, both methods are in agreement with the analytical results within a factor of 6.

\subsection{Biases in Tensor-to-Scalar Ratio}

\begin{figure*} \label{fig:bar_r}
\begin{center}
  $\leavevmode
\begin{array}{c@{\hspace{1cm}}c}
\includegraphics[scale =0.28] {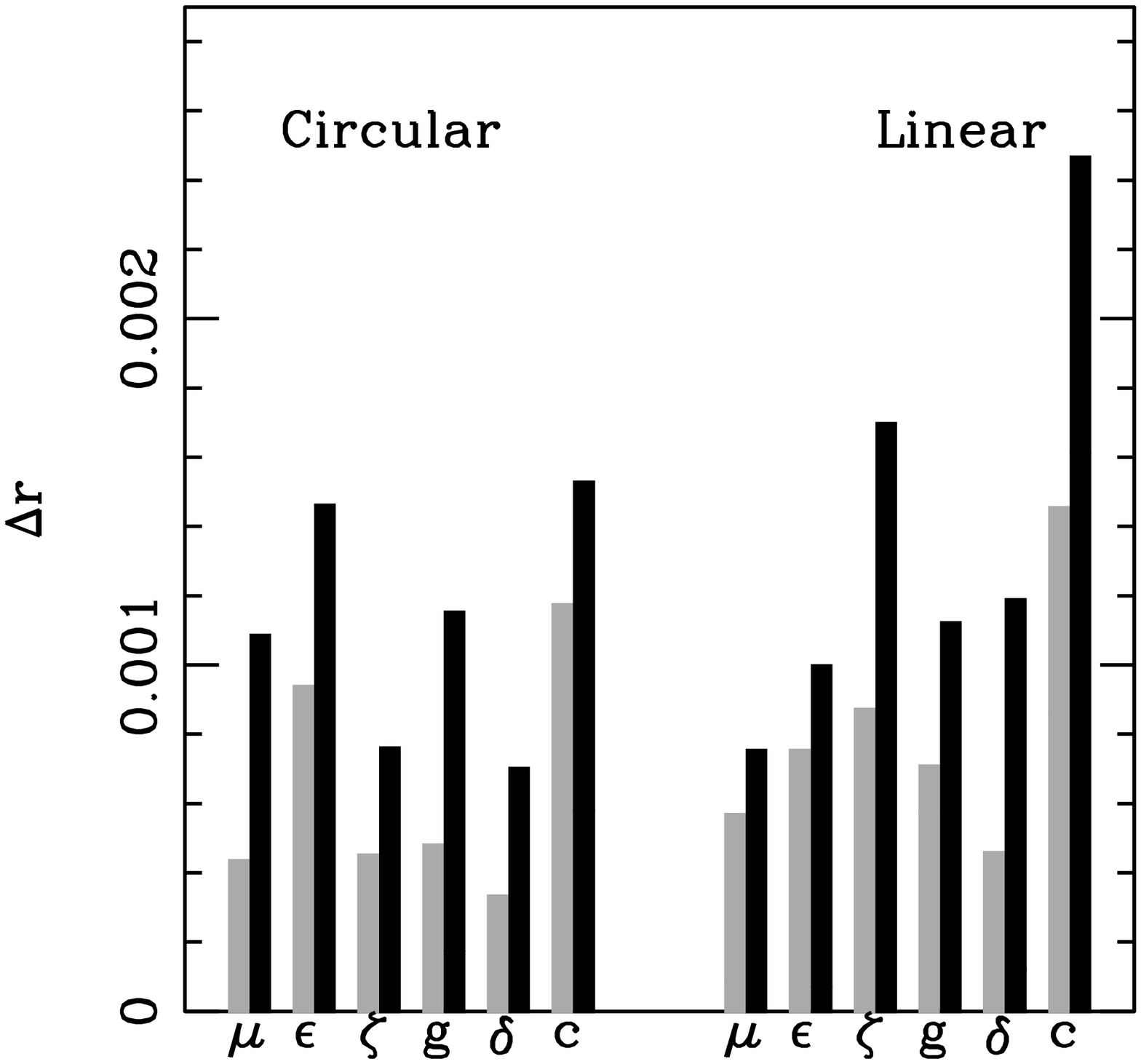} &
\includegraphics[scale =0.28] {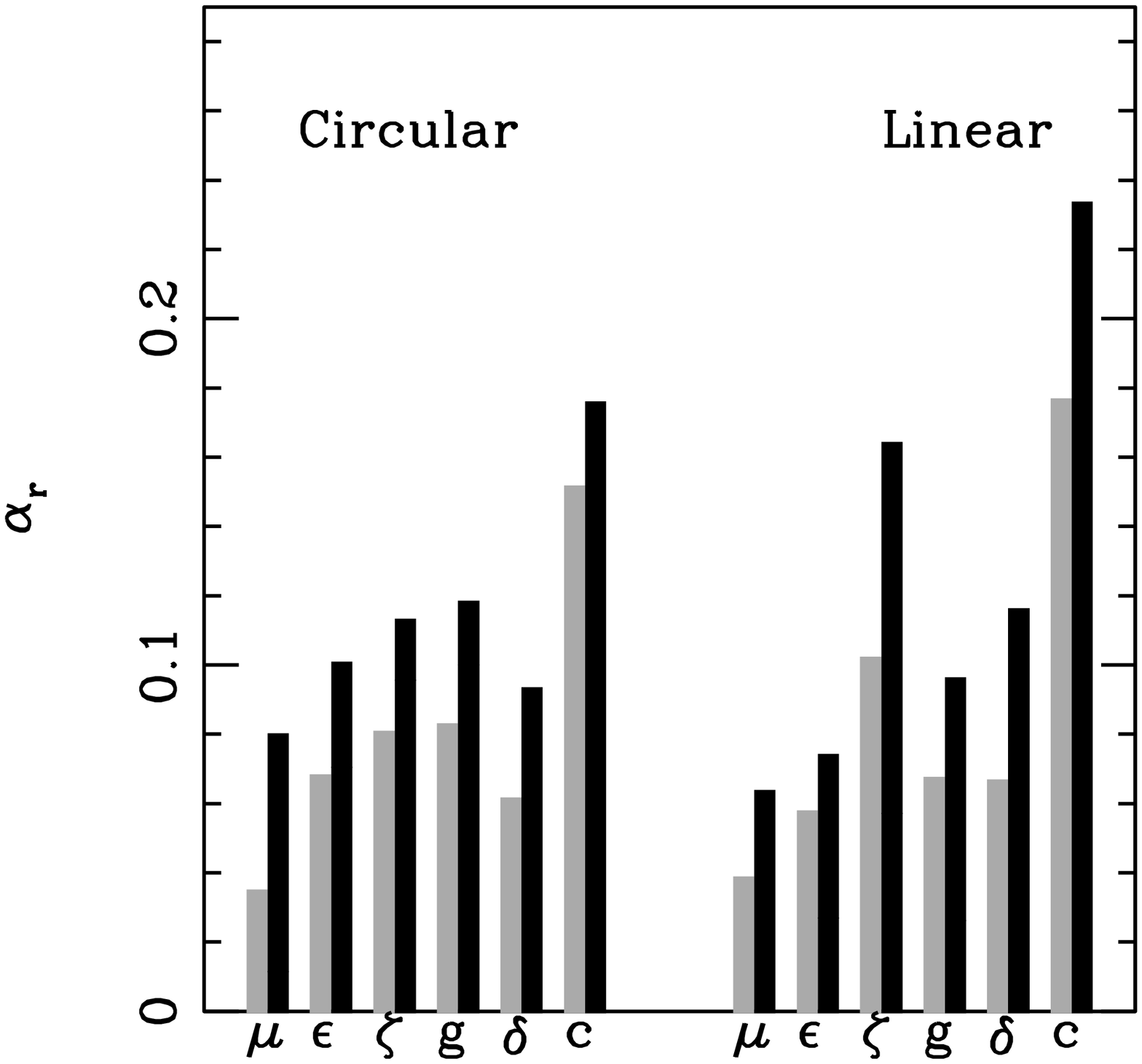} \\
   \end{array}$
\caption{The simulated systematic-induced biases in the tensor-to-scalar ratio $r$ (left) and $\alpha_r$ (right) for the same systematic errors as in Figures~3 and ~4.   All the results derived from the maximum-likelihood (gray) and Gibbs-sampling (black) analyses based on  the simulations in both the linear and circular bases are shown.}
\end{center}
\end{figure*}

The major goal of QUBIC-like experiments is to detect the signals of the primordial B-modes, the magnitude of which is characterized by the tensor-to-scalar ratio $r$. In this context, it is necessary to propagate the effects of systematic errors through to $r$ to assess properly the systematic-induced biases in the primordial B-mode measurements. 

The shape of the primordial $BB$ power spectrum $C^{BB}_{\ell,prim}$ is insensitive to $r$ but the amplitude is directly proportional to $r$. We can straightforwardly convert the amplitude of the systematic-induced false $BB$ into the bias in $r$ by writing $C^{BB}_{\ell,prim} = r~C^{BB}_{\ell, r=1}$ in Eqs.~\ref{eq:rval} and~\ref{eq:alfr} where $C^{BB}_{\ell, r=1}$ is the CAMB~\citep{Lewis:1999bs} calculated primordial $BB$ power spectrum at $r = 1$. The tensor-to-scalar ratios obtained from ideal linear experiment by Gibbs sampling and maximum likelihood methods are found as $r_{GS} = 0.026 \pm 0.012$ and $r_{ML} = 0.006 \pm 0.0095$, respectively. A more conservative estimation for $r$ can be obtained without subtracting the lensed spectrum in Eq.~\ref{eq:rval} and by taking only the first bin where the effect of lensing is the least; $r^{lensed}_{GS} = 0.038 \pm 0.014$ and $r^{lensed}_{ML} = 0.0196 \pm 0.011$.

We vary each systematic error individually and also consider the cross contributions between each error. In realistic observations, all different systematic errors are likely to occur at the same time and we need to understand their combined effects well. We thus evaluate such effects by simulating the systematic errors occurring simultaneously during the observation. The individual and combined systematic-induced biases in $r$ are illustrated in Figure~6, evaluated by both the GS and ML methods based on the simulations performed in the linear and circular bases. Both methods demonstrate good agreement, within a factor of 2.5. Although the mock visibility data are simulated based on only one realization of CMB anisotropy fields, drawn from the power spectra with input $BB$ for $r=0.01$, the resulting false $BB$ band-powers for the different systematic errors  are expected to be a good approximation for other $r$ values since the leading-order false $B$-modes are contaminated only by the leakage of $TT$, $TE$ and $EE$ power spectra, which are independent of $r$. 

The simulations show that, due to the leakage of $TT$ signals into $BB$, even though the cross-polarization and coupling errors are very small, e.g., $\mu_{rms} = 5 \times 10^{-4}$ and $|\epsilon_{rms}| = 5\times 10^{-4}$, the resulting biases in $r$ are comparable to those induced by relatively larger pointing, gain and shape errors. In addition, when increasing the cross-polarization and coupling errors by a factor of 10, the simulations show that the resulting biases would roughly increase by the same factor. As expected, the systematic errors are approximately linearly proportional to their error parameters.  We also find that the combined systematic effects  (referred to as ``c'' in Figure~6) would increase the biases and their values are consistent with the quadrature sum of the individual errors within $10\%$.

If we set up an allowable tolerance level of $10\%$ on $r$, where $r$ is assumed to be $r=0.01$, for QUBIC-like experiments the error parameters adopted as in Figure~6 satisfy this threshold when each systematic error occurs alone during observations. But if all the systematic errors are present at the same time, on average, we require roughly 2 times better systematic control on each error parameter. Although the tolerance level for $r$ is  chosen to be $\alpha_r = 0.1$, our results can directly apply to any other desired threshold level as long as the linear dependence of systematic effects on error parameters is a good approximation for sufficiently small error parameters.

\section{Discussions}

In this work a complete pipeline of simulations is developed to diagnose the effects of systematic errors on the CMB polarization power spectra obtained by an interferometric observation. A realistic, QUBIC-like interferometer design with systematics that incorporate the effects of sky-rotation is simulated. The mock data sets are analyzed by both the maximum likelihood method and the method of Gibbs sampling. The results from both methods are found to be consistent with each other, as well as with the analytical estimations within a factor of 6. 

In order to assess the level at which systematic effects must be controlled, a tolerance level of $\alpha_r = 0.1$ is chosen. This ensures that the instrument is sensitive enough to detect the $B$-signal at $r = 0.01$ level \citep{O'Dea:2006di}. We see that, for a QUBIC-like experiment, the contamination of the tensor-to-scalar ratio at $r = 0.01$ does not exceed the $10\%$ tolerance level in the multipole range $28 < \ell < 384$ when the Gaussian-distributed systematic errors are controlled with precisions of $|g_{rms}| = 0.1$ for antenna gain, $|\epsilon_{rms}| = 5 \times 10^{-4}$ for antenna coupling, $\delta_{rms} \approx 0.7^\circ$ for pointing, $\zeta_{rms} \approx 0.7^\circ$ for beam shape, and $\mu_{rms} = 5 \times 10^{-4}$ for beam cross-polarization when each error acts individually. However, in a realistic experiment all the systematic errors are simultaneously present, in which case the tolerance parameter of $r$ roughly reaches the $20\%$ level, suggesting that better control of systematics would be needed. 

Apart from the systematics presented in the paper, we also ran simulations to analyze the effects of uncertainties in the positions of the antennas. In order to have an effect on the order of $\alpha^{BB} = 0.1$, we found that the uncertainty in the position of each antenna should be on the order of $50\%$ of the length of the $uv$-plane. Since such an error is unrealistically large, we conclude that the effect of antenna position errors on power spectra is negligible in an interferometric observation.

We have shown that a QUBIC-like experiment has fairly manageable systematics, which is essential for the detection of primordial $B$-modes. Since our interferometer design has a large number of redundant baselines (approximately 10 baselines per visibility), as a further improvement, a self-calibration technique can be employed to significantly reduce the level of instrumental errors \citep{Liu:2010yb}.

\section*{Acknowledgments}
Computing resources were provided by the University of Richmond under NSF Grant 0922748. Our implementation of the Gibbs sampling algorithm uses the open-source PETSc library (Balay et al. 1997, 2010) and FFTW \citep{Frigo2005}. G. S. Tucker and A. Karakci acknowledge support from NSF Grant AST-0908844. P. M. Sutter and B. D. Wandelt acknowledge support from NSF Grant AST-0908902. B. D. Wandelt acknowledges funding from an ANR Chaire d'Excellence, the UPMC Chaire Internationale in Theoretical Cosmology, and NSF grants AST-0908902 and AST-0708849. L. Zhang and P. Timbie acknowledge support from NSF Grant AST-0908900. E. F. Bunn acknowledges support from NSF Grant AST-0908900. We are grateful for the generous hospitality of The Ohio State University's Center for Cosmology and Astro-Particle Physics, which hosted a workshop during which some of these results were obtained.
%\vspace{0.9mm}

\bibliography{systErr}

\begin{thebibliography}{51}
\expandafter\ifx\csname natexlab\endcsname\relax\def\natexlab#1{#1}\fi

\bibitem[{{Baker} {et~al.}(1999){Baker}, {Grainge}, {Hobson}, {Jones},
  {Kneissl}, {Lasenby}, {O'Sullivan}, {Pooley}, {Rocha}, {Saunders}, {Scott},
  \& {Waldram}}]{1999MNRAS.308.1173B}
{Baker}, J.~C., {Grainge}, K., {Hobson}, M.~P., {et~al.} 1999, \mnras, 308,
  1173, 1173

\bibitem[{Bond {et~al.}(1998)Bond, Jaffe, \& Knox}]{Bond:1998zw}
Bond, J., Jaffe, A.~H., \& Knox, L. 1998, Phys.Rev., D57, 2117, 2117

\bibitem[{Bunn(2007)}]{Bunn:2006nh}
Bunn, E.~F. 2007, Phys.Rev., D75, 083517, 083517

\bibitem[{Bunn(2011)}]{Bunn:2010kf}
---. 2011, Phys.Rev., D83, 083003, 083003

\bibitem[{{Bunn} \& {White}(1997)}]{1997ApJ...480....6B}
{Bunn}, E.~F., \& {White}, M. 1997, \apj, 480, 6, 6

\bibitem[{{Bunn} \& {White}(2007)}]{2007ApJ...655...21B}
---. 2007, \apj, 655, 21, 21

\bibitem[{{Dickinson} {et~al.}(2004){Dickinson}, {Battye}, {Carreira},
  {Cleary}, {Davies}, {Davis}, {Genova-Santos}, {Grainge}, {Guti{\'e}rrez},
  {Hafez}, {Hobson}, {Jones}, {Kneissl}, {Lancaster}, {Lasenby}, {Leahy},
  {Maisinger}, {{\"O}dman}, {Pooley}, {Rajguru}, {Rebolo}, {Rubi{\~n}o-Martin},
  {Saunders}, {Savage}, {Scaife}, {Scott}, {Slosar}, {Sosa Molina}, {Taylor},
  {Titterington}, {Waldram}, {Watson}, \& {Wilkinson}}]{2004MNRAS.353..732D}
{Dickinson}, C., {Battye}, R.~A., {Carreira}, P., {et~al.} 2004, \mnras, 353,
  732, 732

\bibitem[{{Fomalont} {et~al.}(1984){Fomalont}, {Kellermann}, {Wall}, \&
  {Weistrop}}]{1984Sci...225...23F}
{Fomalont}, E.~B., {Kellermann}, K.~I., {Wall}, J.~V., \& {Weistrop}, D. 1984,
  Science, 225, 23, 23

\bibitem[{Frigo \& Johnson(2005)}]{Frigo2005}
Frigo, M., \& Johnson, S. 2005, Proceedings of the IEEE, 93, 216 , 216

\bibitem[{Gelman \& Rubin(1992)}]{Gelman92}
Gelman, A., \& Rubin, D. 1992, Statistical Science, 7, 457, 457

\bibitem[{{Gorski} {et~al.}(1996){Gorski}, {Banday}, {Bennett}, {Hinshaw},
  {Kogut}, {Smoot}, \& {Wright}}]{1996ApJ...464L..11G}
{Gorski}, K.~M., {Banday}, A.~J., {Bennett}, C.~L., {et~al.} 1996, \apjl, 464,
  L11, L11

\bibitem[{{Grainge} {et~al.}(2003){Grainge}, {Carreira}, {Cleary}, {Davies},
  {Davis}, {Dickinson}, {Genova-Santos}, {Guti{\'e}rrez}, {Hafez}, {Hobson},
  {Jones}, {Kneissl}, {Lancaster}, {Lasenby}, {Leahy}, {Maisinger}, {Pooley},
  {Rebolo}, {Rubi{\~n}o-Martin}, {Sosa Molina}, {{\"O}dman}, {Rusholme},
  {Saunders}, {Savage}, {Scott}, {Slosar}, {Taylor}, {Titterington}, {Waldram},
  {Watson}, \& {Wilkinson}}]{2003MNRAS.341L..23G}
{Grainge}, K., {Carreira}, P., {Cleary}, K., {et~al.} 2003, \mnras, 341, L23,
  L23

\bibitem[{Hinshaw {et~al.}(2012)Hinshaw, Larson, Komatsu, Spergel, Bennett,
  {et~al.}}]{Hinshaw:2012fq}
Hinshaw, G., Larson, D., Komatsu, E., {et~al.} 2012, ArXiv e-prints,
  arXiv:1212.5226

\bibitem[{Hobson \& Maisinger(2002)}]{Hobson:2002zd}
Hobson, M., \& Maisinger, K. 2002, Mon.Not.Roy.Astron.Soc., 334, 569, 569

\bibitem[{{Hobson} \& {Magueijo}(1996)}]{1996MNRAS.283.1133H}
{Hobson}, M.~P., \& {Magueijo}, J. 1996, \mnras, 283, 1133, 1133

\bibitem[{{Hobson} \& {Maisinger}(2002)}]{2002MNRAS.334..569H}
{Hobson}, M.~P., \& {Maisinger}, K. 2002, \mnras, 334, 569, 569

\bibitem[{{Hu} \& {Dodelson}(2002)}]{2002ARA&A..40..171H}
{Hu}, W., \& {Dodelson}, S. 2002, \araa, 40, 171, 171

\bibitem[{Hu {et~al.}(2003)Hu, Hedman, \& Zaldarriaga}]{Hu:2002vu}
Hu, W., Hedman, M.~M., \& Zaldarriaga, M. 2003, Phys.Rev., D67, 043004, 043004

\bibitem[{{Kamionkowski} {et~al.}(1997){Kamionkowski}, {Kosowsky}, \&
  {Stebbins}}]{1997PhRvL..78.2058K}
{Kamionkowski}, M., {Kosowsky}, A., \& {Stebbins}, A. 1997, Physical Review
  Letters, 78, 2058, 2058

\bibitem[{Karakci {et~al.}(2013)Karakci, Sutter, Zhang, Bunn, Korotkov, Timbie,
  Tucker, \& Wandelt}]{Karakci2012}
Karakci, A., Sutter, P.~M., Zhang, L., {et~al.} 2013, Astrophys.J.Suppl., 204,
  8, 8

\bibitem[{{Knoke} {et~al.}(1984){Knoke}, {Partridge}, {Ratner}, \&
  {Shapiro}}]{1984ApJ...284..479K}
{Knoke}, J.~E., {Partridge}, R.~B., {Ratner}, M.~I., \& {Shapiro}, I.~I. 1984,
  \apj, 284, 479, 479

\bibitem[{Komatsu {et~al.}(2011)Komatsu, Smith, Dunkley, Bennett, Gold,
  Hinshaw, Jarosik, Larson, Nolta, Page, Spergel, Halpern, Hill, Kogut, Limon,
  Meyer, Odegard, Tucker, Weiland, Wollack, \& Wright}]{Komatsal2011}
Komatsu, E., Smith, K.~M., Dunkley, J., {et~al.} 2011, The Astrophysical
  Journal Supplement Series, 192, 18, 18

\bibitem[{{Kovac} {et~al.}(2002){Kovac}, {Leitch}, {Pryke}, {Carlstrom},
  {Halverson}, \& {Holzapfel}}]{2002Natur.420..772K}
{Kovac}, J.~M., {Leitch}, E.~M., {Pryke}, C., {et~al.} 2002, Nature, 420, 772,
  772

\bibitem[{Larson {et~al.}(2011)Larson, Dunkley, Hinshaw, Komatsu, Nolta,
  Bennett, Gold, Halpern, Hill, Jarosik, Kogut, Limon, Meyer, Odegard, Page,
  Smith, Spergel, Tucker, Weiland, Wollack, \& Wright}]{Larsonal2011}
Larson, D., Dunkley, J., Hinshaw, G., {et~al.} 2011, The Astrophysical Journal
  Supplement Series, 192, 16, 16

\bibitem[{Larson {et~al.}(2007)Larson, Eriksen, Wandelt, G√≥rski, Huey,
  Jewell, \& O‚ÄôDwyer}]{Larson2007}
Larson, D.~L., Eriksen, H.~K., Wandelt, B.~D., {et~al.} 2007, The Astrophysical
  Journal, 656, 653, 653

\bibitem[{Lewis {et~al.}(2000)Lewis, Challinor, \& Lasenby}]{Lewis:1999bs}
Lewis, A., Challinor, A., \& Lasenby, A. 2000, Astrophys.J., 538, 473, 473

\bibitem[{Liu {et~al.}(2010)Liu, Tegmark, Morrison, Lutomirski, \&
  Zaldarriaga}]{Liu:2010yb}
Liu, A., Tegmark, M., Morrison, S., Lutomirski, A., \& Zaldarriaga, M. 2010,
  Mon.Not.Roy.Astron.Soc., 408, 1029, 1029

\bibitem[{{Martin} {et~al.}(1980){Martin}, {Partridge}, \&
  {Rood}}]{1980ApJ...240L..79M}
{Martin}, H.~M., {Partridge}, R.~B., \& {Rood}, R.~T. 1980, \apjl, 240, L79,
  L79

\bibitem[{Miller {et~al.}(2009)Miller, Shimon, \& Keating}]{Miller:2008zi}
Miller, N., Shimon, M., \& Keating, B. 2009, Phys.Rev., D79, 063008, 063008

\bibitem[{{Myers} {et~al.}(2006){Myers}, {Sievers}, {Bond}, {Contaldi},
  {Mason}, {Pearson}, \& {Readhead}}]{2006NewAR..50..951M}
{Myers}, S.~T., {Sievers}, J.~L., {Bond}, J.~R., {et~al.} 2006, Nature, 50,
  951, 951

\bibitem[{{Myers} {et~al.}(2003){Myers}, {Contaldi}, {Bond}, {Pen}, {Pogosyan},
  {Prunet}, {Sievers}, {Mason}, {Pearson}, {Readhead}, \&
  {Shepherd}}]{2003ApJ...591..575M}
{Myers}, S.~T., {Contaldi}, C.~R., {Bond}, J.~R., {et~al.} 2003, \apj, 591,
  575, 575

\bibitem[{{Ng}(2001)}]{2001PhRvD..63l3001N}
{Ng}, K.-W. 2001, \prd, 63, 123001, 123001

\bibitem[{O'Dea {et~al.}(2007)O'Dea, Challinor, \& Johnson}]{O'Dea:2006di}
O'Dea, D., Challinor, A., \& Johnson, B. 2007, Mon.Not.Roy.Astron.Soc., 376,
  1767, 1767

\bibitem[{{O'Sullivan} {et~al.}(1995){O'Sullivan}, {Yassin}, {Woan}, {Scott},
  {Saunders}, {Robson}, {Pooley}, {Lasenby}, {Kenderdine}, {Jones}, {Hobson},
  \& {Duffett-Smith}}]{1995MNRAS.274..861O}
{O'Sullivan}, C., {Yassin}, G., {Woan}, G., {et~al.} 1995, \mnras, 274, 861,
  861

\bibitem[{Park {et~al.}(2003)Park, Ng, Park, Liu, \& Umetsu}]{Park:2002ka}
Park, C.-G., Ng, K.-W., Park, C., Liu, G.-C., \& Umetsu, K. 2003, Astrophys.J.,
  589, 67, 67

\bibitem[{{Park} {et~al.}(2003){Park}, {Ng}, {Park}, {Liu}, \&
  {Umetsu}}]{2003ApJ...589...67P}
{Park}, C.-G., {Ng}, K.-W., {Park}, C., {Liu}, G.-C., \& {Umetsu}, K. 2003,
  \apj, 589, 67, 67

\bibitem[{{Partridge} {et~al.}(1988){Partridge}, {Nowakowski}, \&
  {Martin}}]{1988Natur.331..146P}
{Partridge}, R.~B., {Nowakowski}, J., \& {Martin}, H.~M. 1988, Nature, 331,
  146, 146

\bibitem[{{Pearson} {et~al.}(2003){Pearson}, {Mason}, {Readhead}, {Shepherd},
  {Sievers}, {Udomprasert}, {Cartwright}, {Farmer}, {Padin}, {Myers}, {Bond},
  {Contaldi}, {Pen}, {Prunet}, {Pogosyan}, {Carlstrom}, {Kovac}, {Leitch},
  {Pryke}, {Halverson}, {Holzapfel}, {Altamirano}, {Bronfman}, {Casassus},
  {May}, \& {Joy}}]{2003ApJ...591..556P}
{Pearson}, T.~J., {Mason}, B.~S., {Readhead}, A.~C.~S., {et~al.} 2003, \apj,
  591, 556, 556

\bibitem[{{Peterson} {et~al.}(2006){Peterson}, {Bandura}, \&
  {Pen}}]{2006astro.ph..6104P}
{Peterson}, J.~B., {Bandura}, K., \& {Pen}, U.~L. 2006, ArXiv Astrophysics
  e-prints, arXiv:astro-ph/0606104

\bibitem[{{Qubic Collaboration} {et~al.}(2011){Qubic Collaboration},
  {Battistelli}, {Ba{\'u}}, {Bennett}, {Berg{\'e}}, {Bernard}, {de Bernardis},
  {Bordier}, {Bounab}, {Br{\'e}elle}, {Bunn}, {Calvo}, {Charlassier}, {Collin},
  {Coppolecchia}, {Cruciani}, {Curran}, {de Petris}, {Dumoulin}, {Gault},
  {Gervasi}, {Ghribi}, {Giard}, {Giordano}, {Giraud-H{\'e}raud}, {Gradziel},
  {Guglielmi}, {Hamilton}, {Haynes}, {Kaplan}, {Korotkov}, {Land{\'e}},
  {Maffei}, {Maiello}, {Malu}, {Marnieros}, {Martino}, {Masi}, {Murphy},
  {Nati}, {O'Sullivan}, {Pajot}, {Passerini}, {Peterzen}, {Piacentini}, {Piat},
  {Piccirillo}, {Pisano}, {Polenta}, {Pr{\^e}le}, {Romano}, {Rosset},
  {Salatino}, {Schillaci}, {Sironi}, {Sordini}, {Spinelli}, {Tartari},
  {Timbie}, {Tucker}, {Vibert}, {Voisin}, {Watson}, {Zannoni}, \& {QUBIC
  Collaboration}}]{2011APh....34..705Q}
{Qubic Collaboration}, {Battistelli}, E., {Ba{\'u}}, A., {et~al.} 2011,
  Astroparticle Physics, 34, 705, 705

\bibitem[{{Scott} {et~al.}(1996){Scott}, {Saunders}, {Pooley}, {O'Sullivan},
  {Lasenby}, {Jones}, {Hobson}, {Duffett-Smith}, \&
  {Baker}}]{1996ApJ...461L...1S}
{Scott}, P.~F., {Saunders}, R., {Pooley}, G., {et~al.} 1996, \apjl, 461, L1, L1

\bibitem[{Shimon {et~al.}(2008)Shimon, Keating, Ponthieu, \&
  Hivon}]{Shimon:2007au}
Shimon, M., Keating, B., Ponthieu, N., \& Hivon, E. 2008, Phys.Rev., D77,
  083003, 083003

\bibitem[{Stuart \& Ord(1987)}]{GVK514310375}
Stuart, A., \& Ord, J. 1987, Kendall's advanced theory of statistics, 5th edn.

\bibitem[{Su {et~al.}(2011)Su, Yadav, Shimon, \& Keating}]{Su:2010wa}
Su, M., Yadav, A.~P., Shimon, M., \& Keating, B.~G. 2011, Phys.Rev., D83,
  103007, 103007

\bibitem[{Sutter {et~al.}(2012)Sutter, Wandelt, \& Malu}]{Sutter:2011uv}
Sutter, P., Wandelt, B.~D., \& Malu, S. 2012, Astrophys.J.Suppl., 202, 9, 9

\bibitem[{{Takahashi} {et~al.}(2010){Takahashi}, {Ade}, {Barkats}, {Battle},
  {Bierman}, {Bock}, {Chiang}, {Dowell}, {Duband}, {Hivon}, {Holzapfel},
  {Hristov}, {Jones}, {Keating}, {Kovac}, {Kuo}, {Lange}, {Leitch}, {Mason},
  {Matsumura}, {Nguyen}, {Ponthieu}, {Pryke}, {Richter}, {Rocha}, \&
  {Yoon}}]{2010ApJ...711.1141T}
{Takahashi}, Y.~D., {Ade}, P.~A.~R., {Barkats}, D., {et~al.} 2010, \apj, 711,
  1141, 1141

\bibitem[{{Timbie} \& {Wilkinson}(1988)}]{1988RScI...59..914T}
{Timbie}, P.~T., \& {Wilkinson}, D.~T. 1988, Review of Scientific Instruments,
  59, 914, 914

\bibitem[{{Timbie} {et~al.}(2006){Timbie}, {Tucker}, {Ade}, {Ali}, {Bierman},
  {Bunn}, {Calderon}, {Gault}, {Hyland}, {Keating}, {Kim}, {Korotkov}, {Malu},
  {Mauskopf}, {Murphy}, {O'Sullivan}, {Piccirillo}, \&
  {Wandelt}}]{2006NewAR..50..999T}
{Timbie}, P.~T., {Tucker}, G.~S., {Ade}, P.~A.~R., {et~al.} 2006, New Astronomy
  Reviews, 50, 999, 999

\bibitem[{{White} {et~al.}(1999){White}, {Carlstrom}, {Dragovan}, \&
  {Holzapfel}}]{1999ApJ...514...12W}
{White}, M., {Carlstrom}, J.~E., {Dragovan}, M., \& {Holzapfel}, W.~L. 1999,
  \apj, 514, 12, 12

\bibitem[{Yadav {et~al.}(2010)Yadav, Su, \& Zaldarriaga}]{Yadav:2009za}
Yadav, A.~P., Su, M., \& Zaldarriaga, M. 2010, Phys.Rev., D81, 063512, 063512

\bibitem[{{Zhang} {et~al.}(2012){Zhang}, {Karakci}, {Sutter}, {Bunn},
  {Korotkov}, {Timbie}, {Tucker}, \& {Wandelt}}]{Zhang2012}
{Zhang}, L., {Karakci}, A., {Sutter}, P.~M., {et~al.} 2012, ArXiv e-prints,
  arXiv:1209.2676

\end{thebibliography}
\nocite{*}

\appendix

Following ~\citet{Bunn:2006nh}, we obtain first order approximations for the $\Delta \hat C^{XY}_{rms}$ given, for a single baseline, in Eq.~\ref{eq:delce}. For a baseline lying on the $x$-axis, the matrices in Eq.~\ref{eq:delce} are given as

$$
\mathbf{N}_{TT} = \left ( \begin{array}{ccc} 1 & 0 & 0 \\ 0 & 0 & 0 \\ 0 & 0 & 0 \end{array} \right ), ~ 
\mathbf{N}_{EE} = \left [ \left(\overline{c^2}\right)^2 - \left(\overline{s^2}\right)^2 \right ]^{-1} \left( \begin{array}{ccc} 0 & 0 & 0 \\ 0 & \overline{c^2} & 0 \\ 0 & 0 & - \overline{s^2} \end{array} \right ), ~ 
\mathbf{N}_{BB} =  \left [ \left(\overline{c^2}\right)^2 - \left(\overline{s^2}\right)^2 \right ]^{-1} \left( \begin{array}{ccc} 0 & 0 & 0 \\ 0 & -\overline{s^2} & 0 \\ 0 & 0 & \overline{c^2} \end{array} \right )
$$ 

$$
\mathbf{N}_{TE} = {1 \over 2\overline{c}} \left( \begin{array}{ccc} 0 & 1 & 0 \\ 1 & 0 & 0  \\ 0 & 0 & 0 \end{array} \right ), ~ 
\mathbf{N}_{TB} = {1 \over 2\overline{c}} \left( \begin{array}{ccc} 0 & 0 & 1 \\ 0 & 0 & 0 \\ 1 & 0 & 0 \end{array} \right ), ~
\mathbf{N}_{EB} = {1 \over 2(\overline{c^2} - \overline{s^2})} \left( \begin{array}{ccc} 0 & 0 & 0 \\ 0 & 0 & 1 \\ 0 & 1 & 0 \end{array} \right ).
$$
The covariance matrix can be written in block-matrix form as
$
{\mathcal{M}}_{w} = \left ( \begin{array}{cc} \mathbf{M}_0 & \mathbf{M}_1 \\ \mathbf{M}^{\dagger}_1 & \mathbf{M}_2 \end{array} \right )
$
where 
$$\mathbf{M}_0 =  \left( \begin{array}{ccc} C^{TT} & C^{TE} \overline{c} & C^{TB} \overline{c}  \\ C^{TE} \overline{c} & C^{EE} \overline{c^2} + C^{BB} \overline{s^2} & C^{EB} (\overline{c^2} - \overline{s^2}) \\ C^{TB} \overline{c}  & C^{EB} (\overline{c^2} - \overline{s^2}) & C^{EE} \overline{s^2} + C^{BB} \overline{c^2} \end{array} \right ).
$$
For a baseline lying in an arbitrary direction, these matrices must be transformed as $\mathbf{M}_0 \to \mathbf{R}^{-1} \mathbf{M}_0 \mathbf{R}$ and $\mathbf{N}_{XY} \to \mathbf{R}^{-1} \mathbf{N}_{XY} \mathbf{R}$, where $\mathbf{R}$ is the rotation matrix given in Eq.~\ref{eq:rotos}. The resulting expression will, then, be averaged over $\theta$. 

\hbox{}

For instrumental errors $\delta \mathbf{v} = \mathbf{E} \cdot \mathbf{v}$, which gives $\mathbf{M}_1 =  \mathbf{M}_0 \cdot \mathbf{E}^{\dagger}$ and $\mathbf{M}_2 = \mathbf{E} \cdot \mathbf{M}_0 \cdot \mathbf{E}^{\dagger}$.

\hbox{~}

\hbox{~}

\chapter{GAIN ERRORS}

$$ g_1 = {1 \over 2} (g_1^i + g_2^i + g_1^{j*} + g_2^{j*}), ~ g_2 = {1 \over 2} (g_1^i - g_2^i + g_1^{j*} - g_2^{j*})$$

\emph{\underline{Linear Basis}}

$$\gamma_1 = {1 \over 2} (g_1^Q + g_1^U), ~ \gamma_2 = {1 \over 2} (g_1^Q - g_1^U), ~ \gamma_3 = {1 \over 2} (g_2^Q + g_2^U), ~$$
$$ \mathbf{E}^{gain}_{linear} = \left ( \begin{array}{ccc} \gamma_1 & \gamma_3 & 0 \\ 0 & \gamma_1 + \gamma_2 & 0 \\ 0 & 0 & \gamma_1 - \gamma_2 \end{array} \right )$$

\hbox{}

$(\Delta \hat C_{rms}^{TT})^{2} = 8 Re\{\gamma_{1}\}^{2}  (C^{TT})^{2}$

\hbox{}

$(\Delta \hat C_{rms}^{TE})^{2} = (6 Re\{\gamma_{1}\}^{2} + {3 \over 4} Re\{\gamma_{2}\}^{2} - {1 \over 4} Im\{\gamma_{2}\}^{2}) (C^{TE})^{2} + (2 Re\{\gamma_1\}^2 +  {1 \over 4} |\gamma_2|^2)C^{TT}C^{EE}$

\hbox{}

$(\Delta \hat C_{rms}^{EE})^{2} = (8 Re\{\gamma_{1}\}^{2} + 4 Re\{\gamma_{2}\}^{2})  (C^{EE})^{2}$

\hbox{}

$(\Delta \hat C_{rms}^{BB})^{2} =  \overline{s^2} |\gamma_{2}|^{2} (C^{EE})^{2} $

\hbox{}

$(\Delta \hat C_{rms}^{TB})^{2} = {1 \over 4} (3 Re\{\gamma_{2}\}^{2} - Im\{\gamma_{2}\}^{2}) (C^{TE})^{2}  + {1 \over 4} ( |\gamma_{2}|^{2} + 8 Re\{\gamma_{1}\}^{2} ) C^{TT} C^{EE} $

\hbox{}

$(\Delta \hat C_{rms}^{EB})^{2} = Re\{\gamma_{2}\}^{2}  (C^{EE})^{2} + (|\gamma_{2}|^{2} + 2 Re\{\gamma_{1}\}^{2}) C^{EE} C^{BB}$

\hbox{} \quad

\emph{\underline{Circular Basis}}

$$\mathbf{E}^{gain}_{circular} = \left ( \begin{array}{ccc} g_1 & 0 & 0 \\ 0 & g_1 &  i g_2 \\ 0 & -i g_2 & g_1 \end{array} \right )$$

\hbox{}

$(\Delta \hat C_{rms}^{TT})^{2} = 8 Re\{g_{1}\}^{2}  (C^{TT})^{2}$

\hbox{}

$(\Delta \hat C_{rms}^{TE})^{2} = 6 Re\{g_{1}\}^{2}(C^{TE})^{2} + 2 Re\{g_1\}^2 C^{TT}C^{EE}$

\hbox{}

$(\Delta \hat C_{rms}^{EE})^{2} = 8 Re\{g_{1}\}^{2} (C^{EE})^{2}$

\hbox{}

$(\Delta \hat C_{rms}^{BB})^{2} =  2 |g_{2}|^{2}C^{EE} ( C^{BB} + \overline{s^2} C^{EE} )$

\hbox{}

$(\Delta \hat C_{rms}^{TB})^{2} = ({3 \over 2} Im\{g_{2}\}^{2} - {1 \over 2} Re\{g_{2}\}^{2}) (C^{TE})^{2}  + {1 \over 2} |g_{2}|^{2} C^{TT} C^{EE} $

\hbox{}

$(\Delta \hat C_{rms}^{EB})^{2} = 2 Im\{g_{2}\}^{2}  (C^{EE})^{2} $

\hbox{} \quad

\chapter{COUPLING ERRORS}

$$ e_1 = {1 \over 2} (e_1^i + e_2^i + e_1^{j*} + e_2^{j*}), ~ e_2 = {1 \over 2} (e_1^i - e_2^i - e_1^{j*} + e_2^{j*})$$

%\hbox{} \quad

\emph {\underline{Linear Basis}}

$$\epsilon_1 = {1 \over 2} (e_1^Q + e_1^U), ~ \epsilon_2 = {1 \over 2} (e_1^Q - e_1^U), ~ \epsilon_3 = {1 \over 2} (e_2^Q + e_2^U), ~ \epsilon_4 = {1 \over 2} (e_2^Q - e_2^U),$$
$$\mathbf{E}^{coupling}_{linear} = \left ( \begin{array}{ccc} 0 & 0 & \epsilon_1 \\ \epsilon_1 + \epsilon_2 & 0 & - \epsilon_3 - \epsilon_4 \\ \epsilon_1 - \epsilon_2 & \epsilon_3 - \epsilon_4 & 0 \end{array} \right )$$

\hbox{}
\hbox{}

$(\Delta \hat C_{rms}^{TT})^{2} = (3 Re\{\epsilon_1\}^2 - Im\{\epsilon_1\}^2) (C^{TE})^{2} + |\epsilon_1|^2 C^{TT}C^{EE}$

\hbox{}

$(\Delta \hat C_{rms}^{TE})^{2} = 2 (Re\{\epsilon_{1}\}^{2} + Re\{\epsilon_{2}\}^{2}) (C^{TT})^{2}$

\hbox{}

$(\Delta \hat C_{rms}^{EE})^{2} = 2 (|\epsilon_{1}|^2 + |\epsilon_{2}|^2 )C^{TT} C^{EE}$

\hbox{}

$(\Delta \hat C_{rms}^{BB})^{2} = 2(|\epsilon_{1}|^2 + |\epsilon_{2}|^2 )C^{TT} C^{BB} + 2 \overline{s^2}(|\epsilon_{1}|^2 + |\epsilon_{2}|^2 )C^{TT} C^{EE}$

\hbox{}

$(\Delta \hat C_{rms}^{TB})^{2} = 2 (Re\{\epsilon_{1}\}^2 + Re\{\epsilon_{2}\}^2) (C^{TT})^{2}$

\hbox{}

$(\Delta \hat C_{rms}^{EB})^{2} = {1 \over 2}(|\epsilon_{1}|^2 + |\epsilon_{2}|^2 )C^{TT} C^{EE} + {1 \over 2}(3 Re\{\epsilon_{1}\}^2 + 3 Re\{\epsilon_{2}\}^2 - Im\{\epsilon_{1}\}^2 - Im\{\epsilon_{2}\}^2) (C^{TE})^{2} $

\hbox{}

\quad

\emph {\underline{Circular Basis}}

$$\mathbf{E}^{coupling}_{circular} = \left ( \begin{array}{ccc} 0 & e_1 & i e_2 \\ e_1 & 0 & 0 \\ i e_2 & 0 & 0 \end{array} \right )$$

\hbox{}

%\hbox{}

$(\Delta \hat C_{rms}^{TT})^{2} = (3 Re\{e_1\}^2 - Im\{e_1\}^2 + 2 Im\{e_2\}^2) (C^{TE})^{2} + (|e_1|^2 + |e_2|^2)C^{TT}C^{EE}$

\hbox{}

$(\Delta \hat C_{rms}^{TE})^{2} = (Re\{e_{1}\}^{2} + Im\{e_{2}\}^{2}) ((C^{TT})^{2} + (C^{TE})^{2} + C^{TT} C^{EE})$

\hbox{}

$(\Delta \hat C_{rms}^{EE})^{2} = (3 Re\{e_{1}\}^2 - Im\{e_{1}\}^2 + 2 Im\{e_{2}\}^2) (C^{TE})^{2} + (|e_{1}|^2 + |e_{2}|^2)C^{TT} C^{EE} $

\hbox{}

$(\Delta \hat C_{rms}^{BB})^{2} = (|e_{1}|^2 + |e_{2}|^2 )C^{TT} ( C^{BB} + \overline{s^2}C^{EE})$

\hbox{}

$(\Delta \hat C_{rms}^{TB})^{2} = (Re\{e_{1}\}^2 + Im\{e_{2}\}^2) (C^{TT})^{2} $

\hbox{}

$(\Delta \hat C_{rms}^{EB})^{2} = {1 \over 4}(|e_{1}|^2 + |e_{2}|^2 )C^{TT} C^{EE} + {1 \over 4}(3 Re\{e_{1}\}^2 + 3 Im\{e_{2}\}^2 - Im\{e_{1}\}^2 - Re\{e_{2}\}^2) (C^{TE})^{2} $

\hbox{} \quad

\hbox{}

\chapter{POINTING ERRORS}

\hbox{}

Defining $\delta \hat {\mathbf{r}}_k$ as the deviation of the $k^{th}$ antenna's pointing center, we can write, to the first order,

$$A^j(\hat {\mathbf {r}})A^k(\hat {\mathbf {r}}) = exp[- (\hat {\mathbf {r}} - \sigma \boldsymbol{\delta})^{2} / 2\sigma^2],$$ 

where $\boldsymbol{\delta} = (\delta \hat {\mathbf {r}}_j +\delta \hat {\mathbf {r}}_k) / 2 \sigma.$

$$\delta V_Z = - i \sigma \int d^2 \mathbf{k} \tilde{Z}(\mathbf{k}) [\tilde{A^2_0}(\mathbf{k} - 2\pi \mathbf{u})]^{*} [ (\mathbf{k} - 2\pi \mathbf{u}) \cdot \boldsymbol{\delta}_Z ] $$

$ \left < V_X \delta V^*_Y \right > = 0$ and $ \left < \delta V_X \delta V^*_Y \right > = {1 \over 2} (\boldsymbol{\delta}_X \cdot \boldsymbol{\delta}_Y) \left < V_X V_Y^* \right >. $ 

\hbox{}

\emph {\underline{Linear Basis}}

$$ \boldsymbol{\delta}_1 = {1 \over 2} (\boldsymbol{\delta}_Q + \boldsymbol{\delta}_U), \quad \boldsymbol{\delta}_2 = {1 \over 2} (\boldsymbol{\delta}_Q - \boldsymbol{\delta}_U)$$

\hbox{}

$(\Delta \hat{C}_{rms}^{TT})^{2} = |\boldsymbol{\delta}_1|^{2}  (C^{TT})^2 $

\hbox{}

$(\Delta \hat{C}_{rms}^{TE})^{2} = {1 \over 2} |\boldsymbol{\delta}_1|^{2} (C^{TE})^2 + {1 \over 8}( 4 |\boldsymbol{\delta}_1|^{2} +  |\boldsymbol{\delta}_2|^{2}) C^{TT}C^{EE}$

\hbox{}

$(\Delta \hat{C}_{rms}^{EE})^{2} =  (  |\boldsymbol{\delta}_1|^{2}  + {1 \over 2} |\boldsymbol{\delta}_2|^{2}) (C^{EE})^2$

\hbox{}

$(\Delta \hat{C}_{rms}^{BB})^{2} =  |\boldsymbol{\delta}_1|^{2} C^{BB} ( C^{BB} +   2 \overline{s^2} C^{EE}) + {1 \over 2} |\boldsymbol{\delta}_2|^{2} C^{EE} ( C^{BB} +  \overline{s^2} C^{EE} )$

\hbox{}

$(\Delta \hat{C}_{rms}^{TB})^{2} = {1 \over 2}  |\boldsymbol{\delta}_1|^{2} C^{TT} ( C^{BB} + \overline{s^2} C^{EE} ) + {1 \over 8} |\boldsymbol{\delta}_2|^{2} C^{TT}C^{EE}$

\hbox{}

$(\Delta \hat{C}_{rms}^{EB})^{2} = {1 \over 2} |\boldsymbol{\delta}_1|^{2} C^{EE} ( C^{BB} + \overline{s^2} C^{EE}) + {1 \over 8} |\boldsymbol{\delta}_2|^{2} (C^{EE})^2 $

\hbox{}

\quad

\emph {\underline{Circular Basis}} \quad $ \boldsymbol{\delta}_2 = 0$

\hbox{} \quad

\chapter{SHAPE ERRORS}

\hbox{} \quad

The product of two elliptic Gaussian beams can be written as a single elliptic Gaussian:

$$A^j(\hat {\mathbf {r}})A^k(\hat {\mathbf {r}}) = exp \left [ - {(x \cos \beta + y \sin \beta)^2 \over 2 (\sigma + \sigma_x)^2} - {(y \cos \beta - x \sin \beta)^2 \over 2 (\sigma + \sigma_y)^2} \right ],$$

where $\beta$ is the angle between the major axis of the resulting ellipse and the $x$-axis.

$$\delta V_Z = - {1 \over \sigma^2} \int d^2 \mathbf{k} \tilde{Z}(\mathbf{k}) [(\widetilde{A^2_0 \Delta_Z})(\mathbf{k} - 2\pi \mathbf{u})]^{*}$$

where $\Delta_Z(x, y) = x^2 (\zeta^Z_x  \cos^2 \beta+ \zeta^Z_y \sin^2 \beta) + y^2 (\zeta^Z_y \cos^2 \beta + \zeta^Z_x \sin^2 \beta) + xy (\zeta^Z_x - \zeta^Z_y)\sin 2\beta$, and $\zeta^Z_{x,y} = \sigma^Z_{x,y} / \sigma.$

\hbox{} 

The only non-vanishing integrals in the covariance matrix are:

%\hbox{} \quad

$$ \int |\tilde {A^2}|^2 = \pi \sigma^2, ~ \int \tilde{A^2} (\widetilde{x^2 A^2 })^* = \int \tilde{A^2} (\widetilde{y^2 A^2 })^* = {1 \over 2} \pi \sigma^4, ~ \int |\widetilde{x^2 A^2 }|^2 = \int |\widetilde{y^2 A^2 }|^2 = {3 \over 4} \pi \sigma^6,$$

$$ \int (\widetilde{x^2 A^2 })(\widetilde{y^2 A^2 })^* = \int |\widetilde{ x y A^2 }|^2 = {1 \over 4} \pi \sigma^6.$$

$$\zeta^Z_1 = {1 \over 2} (\zeta^Z_x + \zeta^Z_y), ~ \zeta^Z_2 = {1 \over 2} (\zeta^Z_x - \zeta^Z_y); \quad \zeta_{i+} = {1 \over 2} (\zeta_i^Q + \zeta_i^U), ~ \zeta_{i-} = {1 \over 2} (\zeta_i^Q - \zeta_i^U) $$

Averaging over $\beta$ we get $ \left < V_X \delta V^*_Y \right > = - \zeta_1^Y \left < V_X V_Y^* \right > $ and  $ \left < \delta V_X \delta V^*_Y \right > = (2 \zeta_1^X \zeta_1^Y + \zeta_2^X \zeta_2^Y ) \left < V_X V^*_Y \right >. $ 

\hbox{} \quad

\emph {\underline{Linear Basis}}

\hbox{}

$(\Delta \hat {C}_{rms}^{TT})^{2} = (10 \zeta_{1+}^2 + 2 \zeta_{2+}^2)(C^{TT})^{2} $

\hbox{}

$ (\Delta \hat{C}_{rms}^{TE})^2 = (7 \zeta_{1+}^2 + {3 \over 4} \zeta_{1-}^2 + \zeta_{2+}^2) (C^{TE})^2 + ( 3 \zeta_{1+}^2 + {1 \over 2} \zeta_{1-}^2 + \zeta_{2+}^2 + {1 \over 4} \zeta_{2-}^2) C^{TT}C^{EE} $

\hbox{}

$ (\Delta \hat{C}_{rms}^{EE})^2 = (10 \zeta_{1+}^2 + 5 \zeta_{1-}^2 + 2 \zeta_{2+}^2 + \zeta_{2-}^2) (C^{EE})^2$

\hbox{}

$ (\Delta \hat{C}_{rms}^{BB})^2 = (2 \zeta_{1-}^2 + \zeta_{2-}^2) \overline{s^2} (C^{EE})^2 + (12 \zeta_{1+}^2 + 12 \zeta_{1-}^2 + 4 \zeta_{2+}^2 + \zeta_{2-}^2 ) \overline{s^2} C^{EE}C^{BB}  + (10 \zeta_{1+}^2 + 5 \zeta_{1-}^2 + 2 \zeta_{2+}^2 + \zeta_{2-}^2) (C^{BB})^2$

\hbox{}

$ (\Delta \hat{C}_{rms}^{TB})^2 = {3 \over 4} \zeta_{1-}^2  (C^{TE})^2 +  {1 \over 4} (2 \zeta_{1-}^2 + \zeta_{2-}^2)  C^{TT} C^{EE} + (3 \zeta_{1+}^2 + {1 \over 2} \zeta_{1-}^2 + \zeta_{2+}^2 + {1 \over 4} \zeta_{2-}^2) \overline{s^2} C^{TT} C^{EE} $

\hbox{}

$ (\Delta \hat{C}_{rms}^{EB})^2 = ({5 \over 4} \zeta_{1-}^2 + {1 \over 4} \zeta_{2-}^2 + \overline{s^2} (3 \zeta_{1+}^2 + \zeta_{2+}^2)) (C^{EE})^2$

\hbox{}

\quad

\emph {\underline{Circular Basis}} \quad $\zeta_{i-} = 0$ 

\hbox{} \quad

\chapter{CROSS-POLARIZATION}

\hbox{}

The only non-vanishing integrals in the covariance matrix are:

\hbox{}

$$\int |\tilde {A^2}|^2 = \pi \sigma^2, ~  \int |\widetilde{A^2 \rho^2 \cos 2\phi}|^2 =  \int |\widetilde{A^2 \rho^2 \sin 2\phi}|^2 = \pi \sigma^6.$$

\hbox{}

\emph {\underline{Linear Basis}}

$$\mu_1 = {1 \over 2} (\mu^Q + \mu^U), \quad \mu_2 = {1 \over 2} (\mu^Q - \mu^U) $$

$$
\delta I = \mu_1 {\rho^2 \over \sigma^2} (Q \cos 2\phi + U \sin 2\phi), \quad \delta Q = (\mu_1 + \mu_2) {\rho^2 \over \sigma^2} I \cos 2\phi, \quad \delta U = (\mu_1 - \mu_2) {\rho^2 \over \sigma^2} I \sin 2\phi
$$

\hbox{}

$(\Delta \hat{C}_{rms}^{TT})^{2} =  2 \mu_1^2 C^{TT}C^{EE}$

\hbox{}

$(\Delta \hat{C}_{rms}^{TE})^{2} = {1\over 2} (\mu_1^2 + \mu_2^2) (C^{TT})^{2} $

\hbox{}

$(\Delta \hat{C}_{rms}^{EE})^{2} = 2  (\mu_1^2 + \mu_2^2) C^{TT} C^{EE} $

\hbox{}

$(\Delta \hat{C}_{rms}^{BB})^{2} = 2  (\mu_1^2 + \mu_2^2) C^{TT} (C^{BB} + \overline{s^2} C^{EE} )$

\hbox{}

$(\Delta \hat{C}_{rms}^{TB})^{2} = {1 \over 2} (\mu_1^2 + \mu_2^2)  (C^{TT})^{2}$

\hbox{}

$(\Delta \hat{C}_{rms}^{EB})^{2} = {1 \over 2} (\mu_1^2 + \mu_2^2) C^{TT} C^{EE} $

\hbox{}

\quad

\emph {\underline{Circular Basis}}

$$\mu_+ = {1 \over 2} (\mu^i + \mu^j), \quad \mu_- = {1 \over 2} (\mu^i - \mu^j) $$ 

$$
\delta I = \mu_+ {\rho^2 \over \sigma^2} Q \sin 2\phi, \quad \delta Q = \mu_+ {\rho^2 \over \sigma^2} I \sin 2\phi + i \mu_- {\rho^2 \over \sigma^2} U \cos 2\phi, \quad \delta U = -i \mu_- {\rho^2 \over \sigma^2} Q \cos 2\phi
$$

\hbox{}

$(\Delta \hat{C}_{rms}^{TT})^{2} =   \mu_+^2 C^{TT}C^{EE} $

\hbox{}

$(\Delta \hat{C}_{rms}^{TE})^{2} = {1\over 4} \mu_+^2 (C^{TT})^{2} $

\hbox{}

$(\Delta \hat{C}_{rms}^{EE})^{2} = ( \mu_+^2 C^{TT} - 2 \mu_-^2 C^{EE} ) C^{EE}$

\hbox{}

$(\Delta \hat{C}_{rms}^{BB})^{2} = (\mu_+^2 C^{TT} - 2 \mu_-^2 C^{EE})(C^{BB} + \overline{s^2} C^{EE} )$

\hbox{}

$(\Delta \hat{C}_{rms}^{TB})^{2} = {1 \over 4}  ( \mu_+^2 C^{TT} - 2 \mu_-^2 C^{EE} ) C^{TT} $

\hbox{}

$(\Delta \hat{C}_{rms}^{EB})^{2} = {1 \over 4} ( \mu_+^2 C^{TT} - 2 \mu_-^2 C^{EE} ) C^{EE} $

\end{document}